\documentclass[aoas, preprint]{imsart}
\RequirePackage{natbib}

\usepackage[english]{babel}
\usepackage[utf8x]{inputenc}
\usepackage[T1]{fontenc}
\usepackage{graphicx}

\usepackage{amsthm,amsmath,natbib}
\RequirePackage[colorlinks,citecolor=blue,urlcolor=blue]{hyperref}

\startlocaldefs
\newcommand{\pois}{\operatorname{Poisson}}
\newcommand{\gamm}{\operatorname{Gamma}}
\newcommand{\discrete}{\operatorname{Count}}
\newcommand{\COV}{\operatorname{Cov}}
\newcommand{\boldeta}{\boldsymbol{\eta}}

\newcommand{\boldalpha}{\boldsymbol{\alpha}}
\newcommand{\boldbeta}{\boldsymbol{\beta}}
\newcommand{\boldm}{\boldsymbol{m}}
\newcommand{\boldX}{\boldsymbol{X}}
\newcommand{\boldZ}{\boldsymbol{Z}}

\endlocaldefs

\begin{document}

\begin{frontmatter}

\title{Estimation of the number of irregular foreigners in Poland using non-linear count regression models}
\runtitle{Estimation of the number of irregular foreigners}

\begin{aug}
  \author{\fnms{Maciej}  \snm{Beręsewicz}\corref{}\ead[label=e1]{maciej.beresewicz@ue.poznan.pl}} %\thanksref{t1}
  \and
  \author{\fnms{Katarzyna} \snm{Pawlukiewicz}\ead[label=e2]{kasia.zadrogaa@gmail.com}}
  
  %\and
  %\author{\fnms{Third}  \snm{Author}%
  %\ead[label=e3]{third@somewhere.com}%
  %\ead[label=u1,url]{http://www.foo.com}}

  %\thankstext{t1}{This study is partially based on Katarzyna Pawlukiewicz's master's thesis entitled \textit{Estimation of the number of irregular immigrants in Poland using hierarchical Gamma-Poisson model} under the supervision of Maciej Beręsewicz.}
  %\thankstext{t2}{First supporter of the project}
  %\thankstext{t3}{Second supporter of the project}

  \runauthor{Beręsewicz and Pawlukiewicz}

  \affiliation{Poznań University of Economics and Business, Statistical Office in Poznań}

  \address{Maciej Beręsewicz,\\
  Poznań University of Economics and Business \\
  Al. Niepodległości 10\\
  61-875 Poznań \\
  Poland\\ 
  \printead{e1} \\
  \and \\
  Statistical Office in Poznań
  }

\address{Katarzyna Zadroga,\\
  Poznań University of Economics and Business \\
  Al. Niepodległości 10\\
  61-875 Poznań\\
  Poland\\ 
    \printead{e2}}
 
\end{aug}

\begin{abstract}

Population size estimation requires access to unit-level data in order to correctly apply capture-recapture methods. Unfortunately, for reasons of confidentiality access to such data may be limited. To overcome this issue we apply and extend the hierarchical Poisson-Gamma model proposed by \citet{zhang2008developing}, which initially was used to estimate the number of irregular foreigners in Norway.

The model is an alternative to the current capture-recapture approach as it does not require linking multiple sources and is solely based on aggregated administrative data that include (1) the number of apprehended irregular foreigners, (2) the number of foreigners who faced criminal charges and (3) the number of foreigners registered in the central population register.  The model explicitly assumes a~relationship between the unauthorized and registered population, which is motivated by the interconnection between these two groups. This makes the estimation conditionally dependent on the size of regular population, provides interpretation with analogy to registered population and makes the estimated parameter more stable over time.

In this paper, we modify the original idea to allow for covariates and flexible count distributions in order to estimate the number of irregular foreigners in Poland in 2019. We also propose a~parametric bootstrap for estimating standard errors of estimates. Based on the extended model we conclude that in as of 31.03.2019 and 30.09.2019 around 15,000 and 20,000 foreigners and were residing in Poland without valid permits. This means that those apprehended by the Polish Border Guard account for around 15-20\% of the total.

\end{abstract}

%\begin{keyword}[class=MSC]
%\kwd[Primary ]{60K35}
%\kwd{60K35}
%\kwd[; secondary ]{60K35}
%\end{keyword}


\end{frontmatter}

\section{Introduction}

The demand for reliable estimates of the number of foreigners residing in a~given country on a~permanent and temporary basis as well as those that are part of the working population is expressed at various levels, including the central government, as well regional and local authorities. Information about the demographic, social and economic characteristics of foreigners is~particularly important for the implementation of population, migration and economic policies. Another important issue is the scale of unregistered / irregular immigration\footnote{In the paper we interchangeably use three terms -- \textit{unregistered, unauthorized, irregular} -- to denote the same group of foreigners who reside in a~given country without a~valid permit.}, i.e.~remaining outside the administrative systems. There is currently no reliable and direct data source that would provide reliable information in this respect. 

Determining the number of foreigners, including unregistered immigrants, is an important methodological challenge for official statistics. First, administrative registers provide information about the \textit{de iure} (registered) population, while statistics are interested in the \textit{de facto} (registered and unregistered) population. Secondly, foreigners constitute a~hard-to-reach population, i.e. one that cannot be easily estimated using traditional statistical methods. This is because there is no available (exhaustive) sampling frame/list and it is difficult to obtain information from individual units . While some characteristics of hard-to-reach populations can be determined by collecting survey data (for example, the selection of units for a~sample can be done using the snowball method and its extension -- \textit{Respondent Driven Sampling}),  the task of estimating the size of such a~population poses a~methodological challenge. 

A~number of appropriate statistical methods for estimating population sizes based on capture-recapture techniques have been proposed in the literature \citep[for a~recent review see][]{bohning2017}. We can categorise these approaches into two groups: the first one includes those based on a~single data source \citep[cf.][]{van2003point, Bohning2009} and the second -- on~at least two data sources \citep[cf.][]{van2012people, coumans2017estimating}. The effective use of these techniques in practice largely depends on the availability of statistical data and is restricted by the need to meet certain assumptions underlying the individual methods. Dual or triple system capture-recapture methods require access to unit-level data (e.g. in order to calculate recapture counts) and are based on certain assumptions, which it may be difficult to meet in practice. 

However, in practice it is often only possible to obtain aggregated data owing to privacy and sensitivity restrictions. For instance, Statistics Poland does not have access to individual data from police or Border Guard records. In such situations one can apply the residual method \citep{passel2007unauthorized, hanson2006illegal}, single-source capture-recapture based on distributional assumptions about count data \citep[cf.][]{bohning2019identity} or models developed for correcting under-reporting as proposed by  \citet{bailey2005modeling, de2017random, stoner2019hierarchical}. The first and most common method applied in economics, for instance by the Pew Research Center \citep{pew2019methodology}, is the residual method, where the size of the unauthorized population is calculated as the difference between the total number of foreigners, non-citizens (e.g. from census data) and that of authorized non-citizens (e.g. from register data). Single source capture-recapture based are more restrictive and is biased in presence heterogeneity and contamination (dependence between captures). To overcome these issue zero-truncated one-inflated distribution was proposed by \citet{godwin2017estimation} and proved to be equivalent with zero-one truncated distributions \citep{bohning2019identity}. The latter method involves joint modelling of the binary indicator or proportion of under-reporting and observed counts. It also requires a~set of strong covariates for each equation and instrumental variables that are connected only with one of these processes.

In this paper we take a~different approach, which was initially proposed by \citet{zhang2008developing} in an unpublished working paper. The model is based solely on aggregated data, with the assumption of a~non-linear relationship between the registered and unregistered population under a~Gamma-Poisson mixed model. The method requires three data sources: 1) observed irregular population (e.g. apprehensions), 2) foreigners listed in police registers (e.g. criminal charges), and 3) known legal population (e.g. from the population register). In the original paper, \citet{zhang2008developing} used the following datasets for Norway: (1) foreigners who did not have a~valid permit for staying in the country, determined on the basis of expulsion requests at the Norwegian Directorate of Immigration (further divided into those who had applied for asylum and those who had not), (2) foreign citizens who faced criminal charges, and (3) foreign-born persons aged 18 and over, registered in the Central Population Register. The main limitation of this method is the fact that some countries (e.g. UK, USA) do not have a~central population register. 

In this study, we critically assess, reuse and extend this model by including demographic covariates and different distributions of counts to estimate the number of of irregular foreigners in Poland. The structure of the paper is as follows. Section \ref{sec-poland} provides a~description of the situation in Poland, basic definitions of concepts referred to in the paper and data sources used for the estimation. Section \ref{sec-model} describes assumptions of the approach proposed by \citet{zhang2008developing}, the model, its critique and extension, including bootstrap MSE estimation. Section \ref{sec-results} offers a~verification of the assumptions given the available data and estimation results. The paper ends with conclusions and discussion. All codes and data used in the paper are available in the supplementary materials \ref{suppA}.

\section{The population of irregular foreigners in Poland}\label{sec-poland}

\subsection{Basic definitions}

The population of unauthorized immigrants is not only hard to reach but also hard to define. To start with, a~foreign-born person can be classified using three characteristics \citet{zhang2008developing}:

\begin{itemize}
    \item entry status: legal or illegal,
    \item residence status: legal, quasi-legal, temporary or illegal,
    \item working status: legal, illegal or no-work.
\end{itemize}

The exact definition of these categories will vary across countries and over time, as a~result of the dynamism and intricacies of immigration laws. In the paper we focus on the \textit{residence} status. 

In the EU context, the term \textit{irregular migrant} refers to~a third-country national present on the territory of a~Schengen State who does not fulfil, or no longer fulfils, the conditions of entry as set out in the Regulation (EU) 2016/399 (Schengen Borders Code) or other conditions for entry, stay or residence in that EU Member State.  

\citet{pew2019}, which provides estimates of the irregular population calculated by applying the residual method, uses the following definition in the EU context: ``Unauthorized immigrants in this report are people living without a~residency permit in their country of residence who are not citizens of any European Union or European Free Trade Association (EFTA) country. The unauthorized population also includes those born in EU-EFTA countries to unauthorized immigrant parents, since most European countries do not have birthright citizenship. Finally, the European unauthorized immigrant population estimate includes asylum seekers with a~pending decision.''

%According to European Commission the term 'irregular' is preferable to 'illegal' migrant because the latter carries a~criminal connotation, entering a~country in an irregular manner, or staying with an irregular status, is not a~criminal offence but an infraction of administrative regulations. Apart from this, juridically and ethically, an act can be legal or illegal but a~person cannot. Thus more and more the term 'migrant in an irregular situation' or 'migrant with irregular status' is preferred.

According to \cite{eurostat2019} in 2019, ``627,900 non-EU citizens were found to be illegally present in the EU-27. This was up 9.7\% compared with one year before (572,200), but down 69.9 \% when compared with the record level of 2015, when that number present stood at 2,085,500''. EU Member States with the largest numbers of non-EU citizens found to be illegally present in 2019 included Germany (133,500), Greece (123,000), France (120,500) and Spain (62,900), which together accounted for 70.1\% of all non-EU citizens found to be illegally present in the EU-27. The corresponding figure for Poland in 2019 was 26,625, compared to 26,547 in 2018. Note that these statistics are based on border guard reports that will be  discussed in the next section.

For administrative purposes, Polish authorities \citep{border2020} use the term \textit{illegal stay}, which is defined as \textit{a stay which does not comply with the legal provisions describing the conditions that foreigners must meet in order to enter and stay in the Republic of Poland }. More specifically, a~person's stay in the Republic of Poland is regarded as illegal when a~foreigner:

\begin{enumerate}
\item does not hold a~valid visa or another valid document entitling them to enter and stay in Poland,
\item has not left the territory of Poland after their period  of stay in the country has expired,
\item has crossed or attempted to cross the border illegally,
\item performs or has performed work illegally,
\item has undertaken business activity in breach of the regulations,
\item does not hold sufficient means of subsistence for the duration of their intended stay in Poland,
\item is a~person identified in an alert issued in the SIS (Schengen Information System) or in the national database for the purposes of refusing entry
\end{enumerate}

If a~foreigner is found to be staying in Poland illegally, an  administrative procedure is initiated whereby the person is obliged to leave the country.

The legality of a~foreigner's stay in Poland can be carried out by representatives of the following agencies: 

\begin{itemize}
    \item officers of the Customs Service,
    \item officers of the Border Guard,
    \item police officers,
    \item authorised employees of the Office for Foreigners,
    \item authorised employees of the Provincial Office. 
\end{itemize}

Currently, there are two institutions that provide information about irregular foreigners -- the Border Guard on a~quarterly basis and the Office for Foreigners within the Ministry of the Interior and Administration on an annual basis. The latter provides information about the number of third country nationals ordered to leave. In this paper we focus on data obtained from the Polish Border Guard, described in the section below.

\subsection{Data sources}

\subsubsection{Polish Border Guard data}

The Polish Border Guard (PBG) reports the number of irregular foreigners according to the actual place of apprehension, which includes: within the country, at airports, at the border with Ukraine, Belarus and Russia separately. In the case of airports or borders, the legal status of foreigners exiting Poland was verified, i.e.  some of them were found to be irregular (e.g. exceeded their period of stay) and were ordered to leave (i.e. this number is reported by the Office for Foreigners). Since these people were already leaving Poland, no apprehension procedure was involved. Consequently, these cases should not be taken into account while estimating the size of the unauthorized population.

Reports prepared by PBG are compiled on a~quarterly basis and are broken down by sex and age. The current reporting suffers from multiple counts of the same individuals, because PBG does not normally remove duplicates from their quarterly statistics. Fortunately, at our request, the data we received from PBG had been deduplicated by accounting for information about re-apprehensions. Currently, PBG can only specify two levels -- first and second or more apprehensions within a~given year. In our study we focus on persons apprehended only once within the country. Table~\ref{tab-sg-basic} presents statistics for the first and second half of 2019. In the first part of 2019 over 11,000 foreigners were found to stay in Poland illegally, with about 3,200 apprehended within the country. These figures increased in the second half: to over 14,000 and 3,500, respectively. The increase can most likely be attributed to those foreigners whose stay permit issued in the first half of the year expired. Note that most illegal foreigners were stopped while leaving Poland at the border with Ukraine, which is the main source country of  non-citizens in Poland. We also note that the number of re-apprehensions is very low and follows a~zero-truncated one-inflated distribution, thus limiting the possibility of applying single-source capture-recapture.

\begin{table}[ht!]
\centering
\caption{The number of irregular foreigners in Poland by place of apprehension and re-apprehension status in 2019}
\label{tab-sg-basic}
\begin{tabular}{llrrrrrr}
  \hline
Half & Same year & Within country & Airports & Ukraine & Russia & Belarus & Total \\ 
  \hline
  I & No & 3,190 & 710 & 6,879 & 106 & 785 & 11,670\\ 
  I & Yes & 29 & 1 & 0 & 0 & 0 & 30\\ 
  II & No & 3,437 & 1,016 & 8,492 & 143 & 1,052 & 14,140\\ 
  II & Yes & 70 & 0 & 0 & 0 & 0 & 70\\ 
  \hline
\end{tabular}
\end{table}

\subsubsection{Police data}

The second data source used in the study is the National Police Information System (Pol. \textit{Krajowy System Informacji Policji}; KSIP) which is the~main police database containing information about individuals suspected of indictable offenses, persons wanted by the police or attempting to hide their identity, and about lost or stolen property. We were given access to police records about registered individuals containing the following classifications: 1) procedural registrations, 2) criminal registrations, 3) searches for missing and wanted individuals and 4) traffic violations.

We obtained anonymised, unit-level data, containing the following variables: a~pseudo-identifier (each person in the KSIP register has a~unique identifier), the quarter in which the registration was made, sex, age calculated at 28 Jan 2020 (date of data compilation), whether or not the person has a~personal id, citizenship, and residence status (unknown, permanent, temporary stay or unregistered). These data are presented in Table \ref{tab-police-basic}. 

According to the police records, 24,571 foreigners were registered in the first half of 2019 and 28,453 in the second half. The increase is mainly due to the higher number of individuals who committed traffic offences, which in turn may result from more intensive police activity during summer holidays and the Christmas period. In both periods,the dataset contains a~similar share of foreigners registered for permanent residence or temporary stay and those unregistered but there are differences in the categories of police registrations. For instance, most procedural and search registrations involved unregistered foreigners, while most traffic violations concerned registered foreigners. In general, the number of foreigners in the police register is higher than that reported by the Border Guard.

\begin{table}[ht!]
\centering
\caption{The number of foreigners in police records by registration type and residence status (registered for temporary stay or permanent residence) in 2019}
\label{tab-police-basic}
\begin{tabular}{lllrrrrr}
  \hline
Half & Registered & Procedural & Search & Traffic & Criminal & Total \\ 
  \hline
  I & Yes & 1,499 & 715 & 9,286 & 10 & 11,510\\ 
  I & No & 4,046 & 6,522 & 2,477 & 16 & 13,061\\ 
  II & Yes & 2,080 & 878 & 11,988 & 6 & 14,952\\ 
  II & No & 4,644 & 5,979 & 2,867 & 11 & 13,501\\ 
   \hline
\end{tabular}
\end{table}

\subsubsection{The registered (legal) population}

A~foreigner, a~citizen of another EU Member State, who stays in Poland for more than 3 months, is obliged to register. Other foreigners are required to register if their stay is longer than 30 days. Since 2018, each foreigner who stays longer than 30 days and has registered, has been automatically assigned a~personal identification number (PESEL), but if registration is not possible (e.g. no permanent place of residence), such a~person can still apply for a~PESEL number. 

The PESEL register is maintained by the Ministry of Digital Affairs\footnote{Throughout the paper we interchangeably use three descriptive terms to refer to foreigners listed in the PESEL register -- the PESEL population, the registered population or the regular population, meaning those who reside in Poland temporarily or permanently.}. 

% It contains information about: 

% \begin{itemize}
%     \item Polish citizens residing in Poland;
%     \item Polish citizens residing outside Poland who have applied for a~Polish identity document;
%     \item foreigners residing in Poland who:
%     \begin{itemize}
%         \item have obtained the right of permanent residence as citizens of a~EU Member State, as national of a~Member State of the European Free Trade Association (EFTA) -- parties to the Agreement on the European Economic Area or as citizens of the Swiss Confederation
%         \item have obtained the right of permanent residence as family members of a~citizen of a~EU Member State, a~citizen of a~European Free Trade Association (EFTA) Member State -- a~party to the Agreement on the European Economic Area or a~citizen of the Swiss Confederation,
%         \item have been granted a~permanent residence permit or an EU long-term residence permit,
%         \item have been granted refugee status, subsidiary protection, asylum,  a~permit for tolerated stay, temporary protection and a~residence permit for humanitarian reasons.
%     \end{itemize}
% \end{itemize}

Table \ref{tab-pesel-basic} presents information about foreigners in the PESEL register (holding a~PESEL id) broken down by registration type: no address in Poland, temporary stay, permanent residence, deregistered `to nowhere', temporary stay expired\footnote{In such cases it is unclear whether such persons have left the country without deregistering or remain in the country without a~valid residence permit, which means they should be included in the irregular population. Without linking the PESEL register with other sources it is not possible to assess the quality of this variable. Note that these figures are over 10 times as high as the number of illegal stays reported by the Border Guard and thus may contain a~significant number of misclassified cases.}, and residence outside Poland. According to the PESEL register, the majority of registered foreigners did not reside in Poland or their period of temporary stay expired. Moreover, in most cases, their stay was temporary. This is mainly due to the requirements that have to be met in order to qualify for permanent residence. In our study we focus on foreigners that have come for a~temporary stay or permanent residence.

\begin{table}[ht!]
\centering
\caption{The number of foreigners in the PESEL register by registration type at quarter ends in 2019}
\label{tab-pesel-basic}
\begin{tabular}{lrrrrrr}
  \hline
As at & No address & Temporary & Permanent & De-registered & Expired  & Outside \\ 
  \hline
31.03 & 81,202 & 242,318 & 56,476 & 16,158 & 124,368 & 332,256 \\ 
  30.06 & 107,545 & 249,154 & 57,656 & 16,246 & 157,476 & 383,283 \\ 
  30.09 & 134,483 & 246,990 & 59,228 & 16,340 & 196,209 & 441,705 \\ 
  31.12 & 160,868 & 252,245 & 60,440 & 16,386 & 225,690 & 496,374 \\ 
   \hline
\end{tabular}
\end{table}

In section \ref{sec-data-for-model} we provide exact information regarding the sub-population of foreigners analysed in our study.

\section{Theoretical properties of the \citet{zhang2008developing} model}\label{sec-model}

\subsection{Model assumptions}

\citet{zhang2008developing} proposed a~model to estimate the number of foreigners at a~given time (i.e. \textit{census night}, \textit{register reference point}), which relies on administrative data \citep[cf.][]{gerritse2016application} but contrasts with most single-source capture-recapture studies, which use data from the whole year or a~specific period to obtain counts for the models. \citet{zhang2008developing} model is based on the assumption that there is a~relationship between the unauthorized and registered population, which is justified below.

Let $M_t$ be the size of the population of unauthorized residents at the time point of interest $t$. Let $N_t$ be the size of the known \textit{reference} (proxy) population at the same time $t$ (e.g. census night; end of the year). We use $N_t$ to denote the number of foreign-born persons over 18 who are registered, i.e. have a~temporary or permanent residence permit. 

$M_t$ should be regarded as a~random variable and $N_t$ as a~known covariate. Let $f(M_t|N_t)$ denote the conditional probabilistic distribution of $M_t$ given $N_t$. The target parameter is the theoretical size of irregular residents, which is defined as the conditional expectation of $M_t$ given $N_t$ with respect to $f(M_t|N_t)$, denoted by

\begin{equation}
    \xi_t = E(M_t|N_t).
\end{equation}

As \citet{zhang2008developing} notes, the theoretical size is defined as the conditional expectation of a~random variable, which makes it possible to get rid of the spurious variation as long as the reference population size is held fixed. The purpose of introducing $N_t$ is two-fold: (a) it serves as an explanatory variable of the irregular size $M_t$, and (b) it provides an interpretation of the irregular size $M_t$ in analogy to $N_t$. In this way, the theoretical size is a~stable measure of the target variable as variation in $M_t$ is linked to that of $N_t$. 

Moreover, since the chosen $N_t$ is not subject to seasonal variation, neither is the theoretical $\xi_t$. In contrast, $M_t$ defined in a~more naturalistic manner can be expected to vary greatly in the course of one year, being perhaps the highest in the summer months, which is another kind of spurious variation.

There is also a~sociological and economic justification for why $M_t$ depends on the regular population $N_t$. Because irregular foreigners do not have regular job opportunities and cannot claim social and health benefits, they need a~network of contacts with registered residents, who are much better off socially and economically. It is hard to imagine a~completely closed community of Ukrainian or Vietnamese irregular residents in Poland. The first one is the largest immigrant group in Poland as a~result of recent migration flows and the second is one of the most stable in terms of size and is confined to a~relatively small area (living mainly in Warsaw and in neighbouring communes). This explains the choice of the reference population -- the registered population aged 18 and over.

%% This explains the choice of the reference population, namely, foreign-born persons with age 18 or over. It seems reasonable to believe that this group should contain most of the direct contact with the irregular residents from the same country of origin. People with foreign roots who were born in Norway can only have contact with the country of origin through their parents or elder relatives, whereas the contact a~foreign-born person has would not cease to exist although the person by now may hold a~Polish citizenship. Finally, it seems plausible that only adults among them can act as dependable resources for the irregular residents 

\subsection{\citet{zhang2008developing} model}

For both the target and the reference populations, let $i=1,...,C$ be the index of the sub-population classified by the country of citizenship and origin, respectively. For simplicity, we drop the $t$ index denoting the reference time. \citet{zhang2008developing} assumed that the observed number of irregular residents follows a~Poisson distribution, with parameter $\lambda_i$, denoted by 

\begin{equation}
    m_{i} \sim \pois \left( \lambda _ { i } \right).
    \label{eq-poisson-start}
\end{equation}

The parameter $\lambda_i$ should depend on two other quantities: (a) the total number of irregular residents from country $i$, denoted by $M_i$, and  (b) the probability of being observed, i.e. the probability for an irregular resident to be included in Border Guard data, denoted by $p_i$, i.e. $\lambda_i = M_ip_i$. 

In addition, let  $u_i = M_ip_i / E(M_ip_i | n_i, N_i)$, where $E(M_ip_i | n_i, M_i)$ denotes the conditional expectation of $M_ip_i$ given $n_i$ and $N_i$. The  $u_i$ is a~random effect that accounts for heterogeneous variation from one country to another. Together, we obtain

\begin{equation}
    \lambda _ { i } = \mu _ { i } u _ { i }, 
\end{equation}

\noindent where $\mu_{ i } = E \left( M _ { i } p _ { i } | n _ { i } , N _ { i } \right) = E \left( M _ { i } | N _ { i } \right) \cdot E \left( p _ { i } | M _ { i } , n _ { i } , N _ { i } \right)$. The final model is specified by the following set of equations

\begin{eqnarray}
%\begin{aligned}
    \xi _ { i } &=& E \left( M _ { i } | N _ { i } \right) = N _ { i } ^ { \alpha }, \\
    \omega_i & = & E \left( p _ { i } | M _ { i } , n _ { i } , N _ { i } \right) = E \left( p _ { i } | n _ { i } , N _ { i } \right) = \left( \frac { n _ { i } } { N _ { i } } \right) ^ { \beta },\\ 
    u _ { i } &\sim & \gamm ( 1 , \phi ),
%\end{aligned}
\label{eq-zhang-model}
\end{eqnarray}

\noindent where $\gamm ( 1 , \phi )$ denotes the gamma distribution with the expectation $E(u_i)=1$ and variance $V(u_i)=1/\phi$. Zhang uses term hierarchical Gamma-Poisson random effect model to describe \ref{eq-zhang-model} and derives log-likelihood function that may be found in Appendix \ref{appen-ll-der} but we show in Appendix \ref{appen-estimation-mixture} that it is actually Negative Binomial distribution as a special case of Poisson-Gamma mixture. Furthermore, we show in Appendix \ref{appen-stirling} that the simplified approximation of term $\log\Gamma(x)$ used by \citet{zhang2008developing} leads to biased estimates of $\alpha, \beta$ and $\xi$.

The model has the following assumptions. First, country random variation refers only to observed apprehensions i.e. $m_i$ as $\mu_i$ is scaled by $u_i$. Second, the non-linear relationship between the regular population size $N_i$ and $M_i$ is imposed by the power function with $\alpha$ being the same for all the countries. Finally, there is a~similar relationship, defined by the $\beta$ parameter, which exists between police ``catch rate'' and the probability of being observed in Border Guard data.

From equations \eqref{eq-poisson-start}-\eqref{eq-zhang-model} we derive the following relationship 

\begin{equation}
    \mu_i = N_i^\alpha \left( \frac{n_i}{N_i} \right)^\beta,
    \label{eq-mu}
\end{equation}

\noindent that can be used to verify the model assumptions. After dividing both sides by $N_i$ and applying the log transformation we get

\begin{equation}
    \log \left(\frac{\mu_i}{N_i}\right) = (\alpha - 1) \log N_i + \beta \log \left(\frac{n_i}{N_i}\right).
\end{equation}

Then, we can plug in $m_i$ and model $\log \left(m_i / N_i\right)$ into the linearized model

\begin{equation}
    \log \left(\frac{m_i}{N_i}\right) = (\alpha - 1) \log N_i + \beta \log \left(\frac{n_i}{N_i}\right) + \epsilon_i, 
    \label{eq-lin}
\end{equation}

\noindent from which we should expect a~negative relationship with $\log N_i$ and a~positive one with $\log(n_i/N_i)$.

\subsubsection{The target parameters}

We are interested in the target parameter describing the population size of irregular residents. Given the above model, the target parameter is defined as 

\begin{equation}
  \xi = \sum _ { i = 1 } ^ { C } E \left( M _ { i } | N _ { i } \right) = \sum _ { i=1 }^{C} N _ { i } ^ { \alpha },
\end{equation}

\noindent and its estimator is given by 

\begin{equation}
\hat { \xi } = \sum _ { i=1 }^C N _ { i } ^ { \hat { \alpha } },
\label{xi-estimator}
\end{equation}

\noindent where $\hat{\alpha}$ is the estimator of $\alpha$. 

% Another quantity that we are interested in is the average probability of being observed defined as 

% \begin{equation}
%     \overline{\omega} = \frac{1}{C}\sum_{i=1}^C E(p_i|M_i, n_i, N_i) = \frac{1}{C}\sum_{i=1}^C E(p_i|n_i, N_i) = \frac{1}{C} \sum_{i=1}^C\left(\frac{n_i}{N_i} \right)^\beta,
% \end{equation}

% \noindent with its estimator given by

% \begin{equation}
%     \widehat{\overline{\omega}} = \frac{1}{C} \sum_{i=1}^C\left(\frac{n_i}{N_i} \right)^{\hat{\beta}},
% \end{equation}

% \noindent where $\hat{\beta}$ is the estimator of $\beta$. 

\section{Extensions}

A natural way of extending the above model is to include additional covariates, such as country, sex, age or place of residence to account for variability in $\alpha$ or/and $\beta$ and including different distributions for the observed counts $m_{tij}$. 

Let $m_{tij}$ be observed counts for period $i=1,..,T$ , country $i=1,...,C$ and domain $j=1,...,J$ defined as an interaction between, e.g. sex and age. We assume that the observed counts are generated from the count distribution given \eqref{eq-count-general}

\begin{equation}
     m_{tij} \sim  \discrete(\boldeta),
     \label{eq-count-general}
\end{equation}

\noindent where $\discrete$ denotes a~suitable count distribution, such as Poisson, Geometric or Negative Binomial (NB2), and $\boldeta$ is a~vector of parameters for a~given distribution -- for Poisson: $\boldeta=(\mu_{tij})$ or for NB2: $\boldeta=(\mu_{tij}, \phi)$, where $\mu_{tij}$ is defined as in \eqref{eq-mu}.  

Note that, by definition, $\mu_{tij} > 0$ is positive as we only observe apprehended foreigners from a~given country and belonging to a~given domain. This resembles the situation in single-source capture-recapture studies based on re-apprehensions. Thus assuming \eqref{eq-poisson-start} will lead to underestimation estimates of the population total as shown in a limited simulation study in Appendix \ref{appen-stirling}. This results suggest that the original model proposed by \citet{zhang2008developing} may lead to biased estimates of $\xi$. Furthermore, zero-inflation may be presence. For instance, the study by \citet{bohning2019identity} shows equivalence between zero-truncated one-inflated and zero-one truncated count distributions. Having that in mind, our extension we also consider distributions that may be zero-truncated as the one below

\begin{equation}
f_{+}(m_{tij}, \boldeta)= \frac{f(m_{tij}, \boldeta)} {1-p(0, \boldeta)},
\end{equation}

\noindent or zero-one truncated as given by 

\begin{equation}
f_{++}(m_{tij}, \boldeta)= \frac{f(m_{tij}, \boldeta)} {1-f(0, \boldeta)-f(1, \boldeta)},
\end{equation}

\noindent where $f_{+}(.), f_{++}(.)$ denote truncated count densities, $f(0, \boldeta), f(1, \boldeta)$ represent 0 and 1 densities and $\boldeta$ is a~vector of parameters for a~given count distribution. 

% For instance, the density for the Poisson-Gamma distribution is given by equation \eqref{eq-pois-gamma-density} and the truncated distribution will be given by

% \begin{equation}
%     f_{++}\left( m _ { tij } ; \boldeta \right) = 
%     \frac{\frac { \mu _ { tij } ^ { m _ { tij } } \phi ^ { \phi } } { m _ { tij } ! \Gamma ( \phi ) } \left( \mu _ { tij } + \phi \right) ^ { - \left( m _ { tij } + \phi \right) } \Gamma \left( m _ { tij } + \phi \right)}{
%     1 - \frac { \phi ^ { \phi } } { \Gamma ( \phi ) } \left( \phi \right) ^ { - \phi } \Gamma \left( \phi \right)}.
% \end{equation}

In our study we consider the following distributions for $m_{tij}$: Poisson (PO), zero-truncated Poisson (ztPO),  Negative-Binomial (NB2) and zero-truncated Negative Binomial (ztNB2), where $\mu$ is the mean and $\phi$ is the dispersion parameter. Log-likelihood functions for the models considered in the paper are given in Appendix \ref{appen-ll-der}. 

%We did not consider zero-one truncated distributions, which we intend to analyse in future studies. 

Furthermore, we extend equation \eqref{eq-mu} by including covariates 

\begin{equation}
    \mu_{tij} =  N _ { tij } ^ { \boldX^T\boldalpha } \left( \frac { n _ { tij } } { N _ { tij } } \right) ^ { \boldZ^T\boldbeta },
\label{eq-extended-model}
\end{equation}

\noindent where $\boldX$ and $\boldZ$ may be the same and $\boldeta$ is estimated using the maximum likelihood method. Note that $\boldX$ and $\boldZ$  can refer to the domains defined by $j$, as the model uses only two covariates -- population size $N_{tij}$ and police records to population size $n_{tij}/N_{tij}$. This is an interesting alternative to, for instance, the model proposed by \citet{stoner2019hierarchical}, which requires strong covariates for under-reporting and observed counts. 

% To account for variability in $\alpha$ and $\beta$, we used two indicator variables: one for sex (Males = 1, $\boldX_{1}$) and one for Ukraine (the largest group of foreigners). Thus, the full version of \eqref{eq-extended-model} is as follows

% \begin{equation}
%     \mu_{tij} =  N _ { tij } ^ { \alpha_0 + \alpha_1 \boldX_1 + \alpha_2 \boldX_2} \left( \frac { n _ { tij } } { N _ { tij } } \right) ^ { \beta_0 + \beta_1 \boldX_1 + \beta_2 \boldX_2}.
% \label{eq-extended-model-simple}
% \end{equation}

Under the model \eqref{eq-extended-model} estimator for the target parameter given by \eqref{xi-estimator} changes to 

\begin{equation}
\hat { \xi } = \sum _ { i=1 }^C N _ { i } ^ { \boldX^T\hat{\boldalpha} }.
\end{equation}

\section{Estimating uncertainty}

\subsection{\cite{zhang2008developing} proposal}

In the original paper, Zhang did not calculate variance for the target parameter $\xi$ but proposed a~confidence interval for $\xi$ by plugging in the confidence interval for $\alpha$. Thus, the CI for $\xi$ is given by

\begin{equation}
    \left(\sum_{i=1}^C N_i^{\alpha_l}, \sum_{i=1}^C N_i^{\alpha_u}\right),
\end{equation}

\noindent where $\alpha_l,\alpha_u$ are the lower bound and upper bound of the CI interval for $\alpha$. If we use additional covariates to explain variability in $\xi$ we can plug in lower and upper bounds for all parameters as shown below

\begin{equation}
    \left(\sum_{i=1}^C N_i^{\boldX^T\boldalpha_l}, \sum_{i=1}^C N_i^{\boldX^T\boldalpha_u}\right),
\end{equation}

\noindent where $\boldalpha_l$ and $\boldalpha_u$ are vectors with lower and upper bounds.

\subsection{Parametric bootstrap} 

We use an alternative approach, based on parametric bootstrapping, to estimate the mean square error, which exploits an idea similar to that proposed by \citet{gonzalez2008bootstrap} for small area estimation. It consisting of the following steps: 

\begin{enumerate}
    \item given $n_{i}, m_{i}$ and $N_{i}$, calculate $\hat{\boldeta}$ using the maximum likelihood function,
    \item given $\hat{\boldeta}$, generate $\hat{\boldeta}^*$ from a multivariate normal distribution $\mathrm{MVN}(\hat{\boldeta}, \widehat{\COV}(\hat{\boldeta}))$, where $\widehat{\COV}$ denotes the covariance of $\hat{\boldeta}$. For instance, for the NB2 model we use
    
\begin{equation}
\left[
\begin{array}{c}
\boldalpha^* \\ 
\boldbeta^* \\
\phi^*
\end{array}
\right]
\sim 
\mathrm{MVN}
\left(
\left[
\begin{array}{c}
\hat{\boldalpha} \\  \hat{\boldbeta} \\ \hat{\phi}
\end{array}
\right],
\left[
\begin{array}{ccc}
%% first row 
V(\hat{\boldalpha}) & 
\COV(\hat{\boldalpha}, \hat{\boldbeta}) &  
\COV(\hat{\boldalpha}, \hat{\phi}) \\
%% second row 
\COV(\hat{\boldbeta},\hat{\boldalpha}) &
V(\hat{\boldbeta})  &  
\COV(\hat{\boldbeta}, \hat{\phi}) \\ 
\COV(\hat{\phi}, \hat{\boldalpha}) & 
\COV(\hat{\phi}, \hat{\boldbeta}) &  
V(\hat{\phi})
\end{array}\right]
%% third row 
\right),
\end{equation}
    \item calculate $\xi^* = \sum_{i=1}^C N_i^{\boldalpha^*}$,
    \item generate $m_i^*$ from the assumed distribution using $\mu^*=N^{\boldX^T\boldalpha^*}\left(\frac{n}{N}\right)^{\boldZ^T\boldbeta^*}$,
    \item fit the model to $(m_i^*, n_{tij}, N_{tij})$ and estimate $\boldeta^*$,  
    \item estimate $\hat{\xi}^* = \sum_{i=1}^C N_i^{\hat{\boldalpha}^*}$,
    \item repeat steps 2--6 $B$ times and calculate the bootstrap MSE estimator
    \begin{equation}
        \operatorname{mse} = \frac{1}{B}\sum_{b=1}^B\left(\hat{\xi}^*-\xi^*\right)^2,
    \end{equation}
    and Relative MSE estimator
     \begin{equation}
        \operatorname{rmse} = 
        \frac{\sqrt{\operatorname{mse}}}{\bar{\xi}^*}.
    \end{equation}
    \end{enumerate}
    
    Based on bootstrapped $\xi^*$ from point 4 we calculate the confidence interval for $\xi$ using the 95\% percentile method and a method recently introduced by \citet{liu2015simulation} called the \textit{shortest probability interval} (SPIN). The latter method is recommended for asymmetric distributions, bounded variables (e.g. positive); intervals constructed using SPIN have better coverage.

\section{Results}\label{sec-results}

\subsection{Data for the model}\label{sec-data-for-model}

In our study we used Polish data from two halves of 2019 for the foreign population aged 18+. The PESEL register reflected the state at 31 March and 30 September. Then, we prepared data for the first and second half of the year using police and Border Guard data. \citet{zhang2008developing} used a~similar approach involving population data as at 01 Jan 2006, police data about foreigners charged with criminal offences in 2005 and the number of unauthorized foreigners between May 2005 and April 2006. In addition, we derived data broken down by sex and economic age group (18-59 and 60+ for women; 18-64 and 65+ for men).  Table \ref{tab-model-data}  presents information about the number of foreigners and countries of origin present in the PESEL, police and Border Guard registers. The PESEL register contained 151 and 147 countries in the first and second half of the year, respectively, police data -- around 100, and Border Guard records -- around 70. The two latter sources contain a~considerably greater percentage of men, in contrast to the PESEL register, where women account for around 60\% of all foreigners.

\begin{table}[ht!]
\centering
\caption{The number of foreigners and countries by data source, sex and period before applying the condition for the model}
\label{tab-model-data}
\begin{tabular}{llrrrr}
  \hline
   \multicolumn{2}{c}{Classification} & 
   \multicolumn{2}{c}{Number of foreigners} & 
   \multicolumn{2}{c}{Number of countries} \\
  Source & Sex & 1$^{st}$ period & 2$^{st}$ period & 1$^{st}$ period &  2$^{st}$ period\\
  \hline
  PESEL & Total & 232,468 & 234,194 & 151 & 147 \\
        & Women& 137,424 & 137,880	 & 145 & 140 \\
        & Men & 95,044 & 96,314 & 127 & 130  \\
  Border Guard & Total & 3,187 & 3,435 & 77 & 68 \\ 
                & Women & 762 & 776 & 40 & 39 \\ 
                & Men & 2,425 & 2,659  & 72 & 67 \\         
  Police (all) & Total & 20,138 & 23,330 & 100 & 98 \\ 
                & Women & 3,017 & 3,079 & 58 & 57 \\ 
                & Men & 17,121 & 20,251 & 94 & 94 \\ 
   \hline
\end{tabular}
\end{table}

The model requires that the following conditions hold: $m_{tij} >0$, $n_{tij} >0 $ and $n_{tij}/N_{tij} < 1$, so we created a~new dataset that meets these requirements. Countries that do not satisfy these conditions were grouped to create a~pseudo-country denoted as \textit{other}\footnote{In the plots they are marked as UNK, i.e. unknown}. After applying this condition, we received a~total of 73 countries (including category \textit{other}), of which 50 were observed in both periods and 23 only in one (65 in the first and 58 in the second half of 2019). The full list of countries is given in Appendix \ref{appen-data}.

% \begin{table}[ht!]
% \centering
% \caption{Data used for the model -- number of foreigners by period, sex and data source}
% \label{tab-model-data}
% \begin{tabular}{llrrr}
%   \hline
% Half & Sex & PESEL & Border & Police \\ 
%   \hline
% I half & Females & 137,424 & 762 & 3,017 \\ 
%   & Males & 95,044 & 2,425 & 17,121 \\ 
%       & Total & 232,468 & 3,187 & 20,138 \\ 
%   II half & Females & 137,880 & 776 & 3,079 \\ 
%     & Males & 96,314 & 2,659 & 20,251 \\ 
%      & Total & 234,194 & 3,435 & 23,330 \\ 
%   \hline
% \end{tabular}
% \end{table}

\subsection{Verification of the assumptions}

\begin{figure}[ht!]
    \centering
    \includegraphics[width=0.8\textwidth]{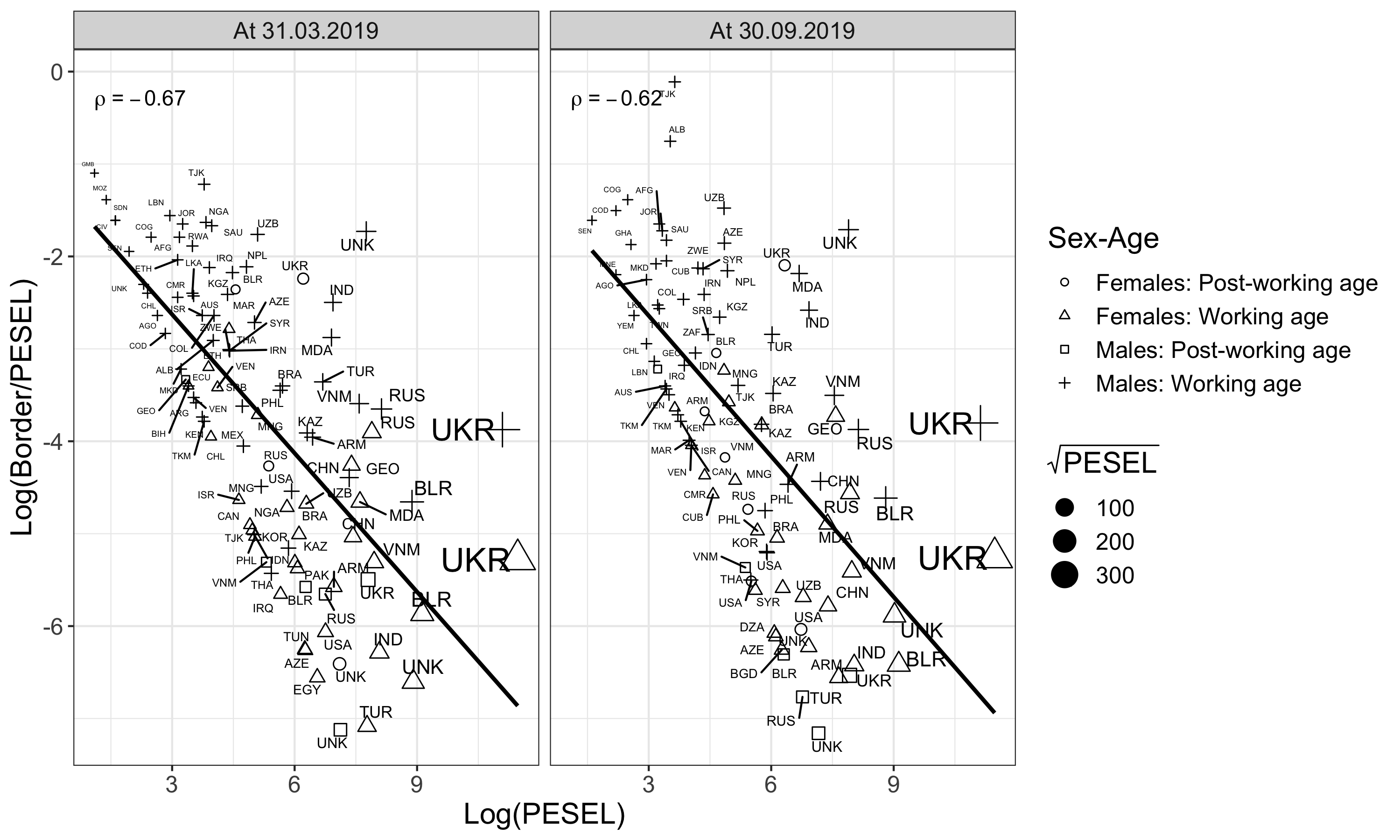}
    \includegraphics[width=0.8\textwidth]{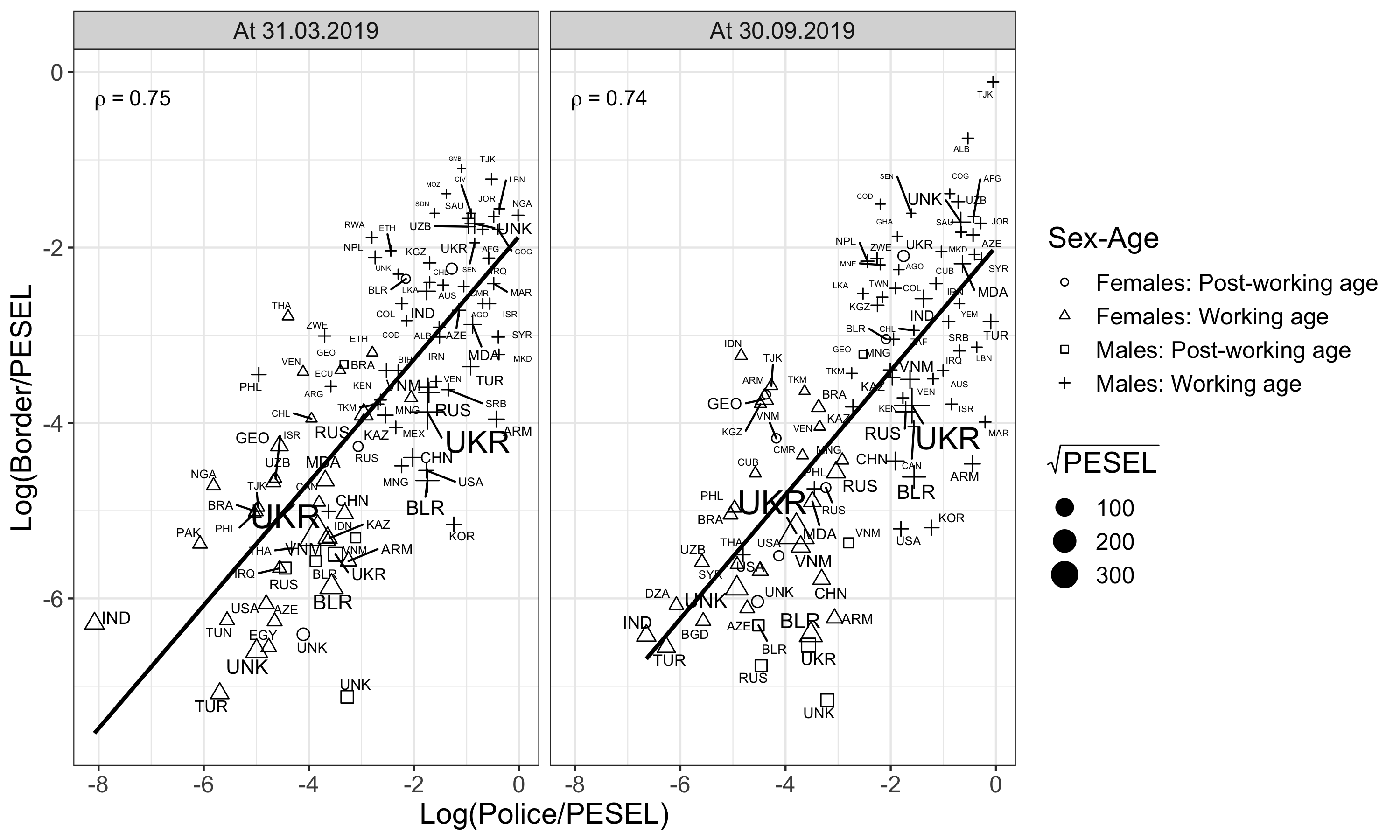}
    \caption{The relationship between the log of the PESEL population and the log of the BG-to-PESEL counts (top) and between the log of police-to-PESEL counts and the log of BG-to-PESEL counts (bottom) at the end of first and third quarter of 2019. Shapes represent domains cross-classified by sex and age, symbol size represents the square root of the PESEL population and solid lines are regression lines. Pearson correlation coefficient is denoted by $\rho$ in the top left corner.}
    \label{fig-overall}
\end{figure}

\begin{figure}[ht!]
    \centering
    \includegraphics[width=0.8\textwidth]{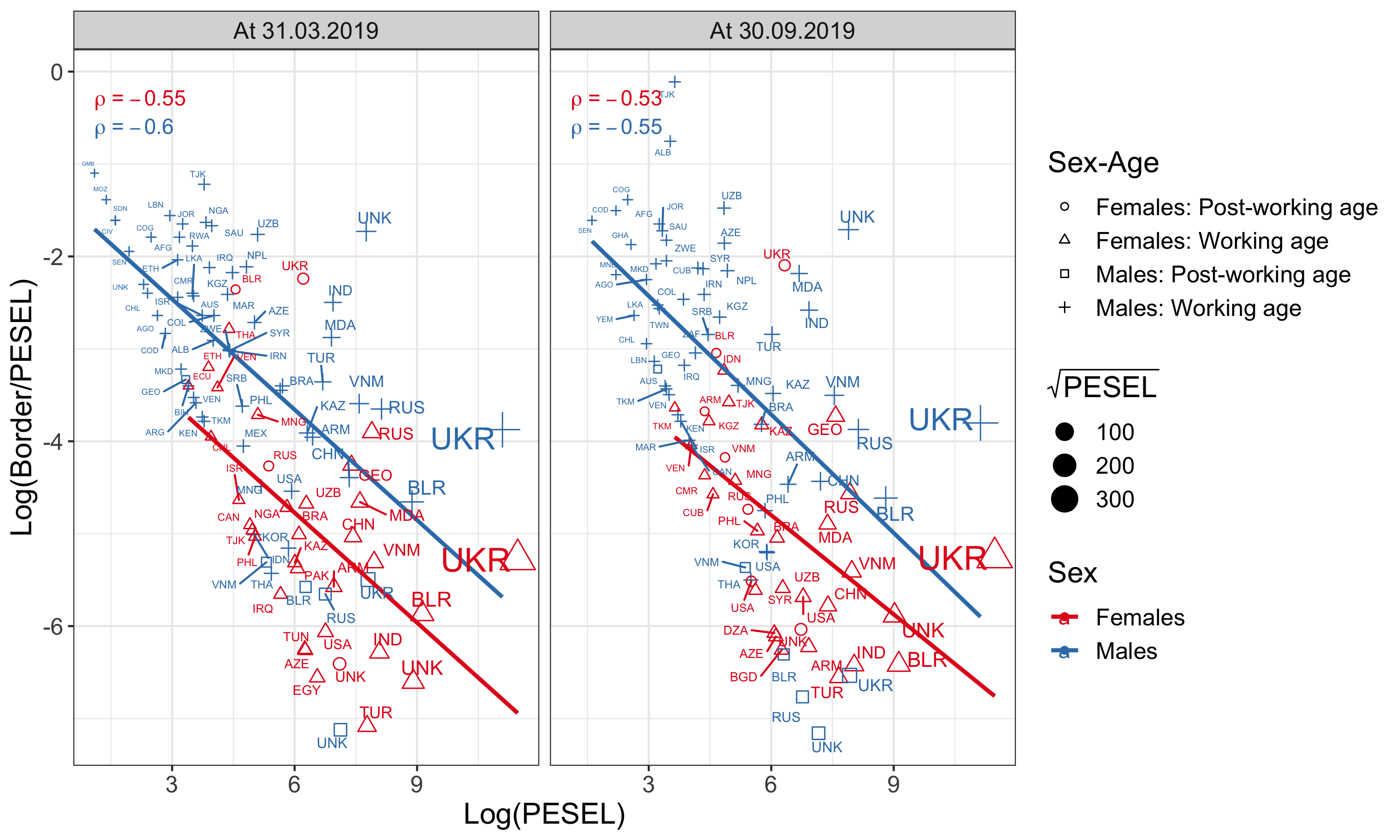}
    \includegraphics[width=0.8\textwidth]{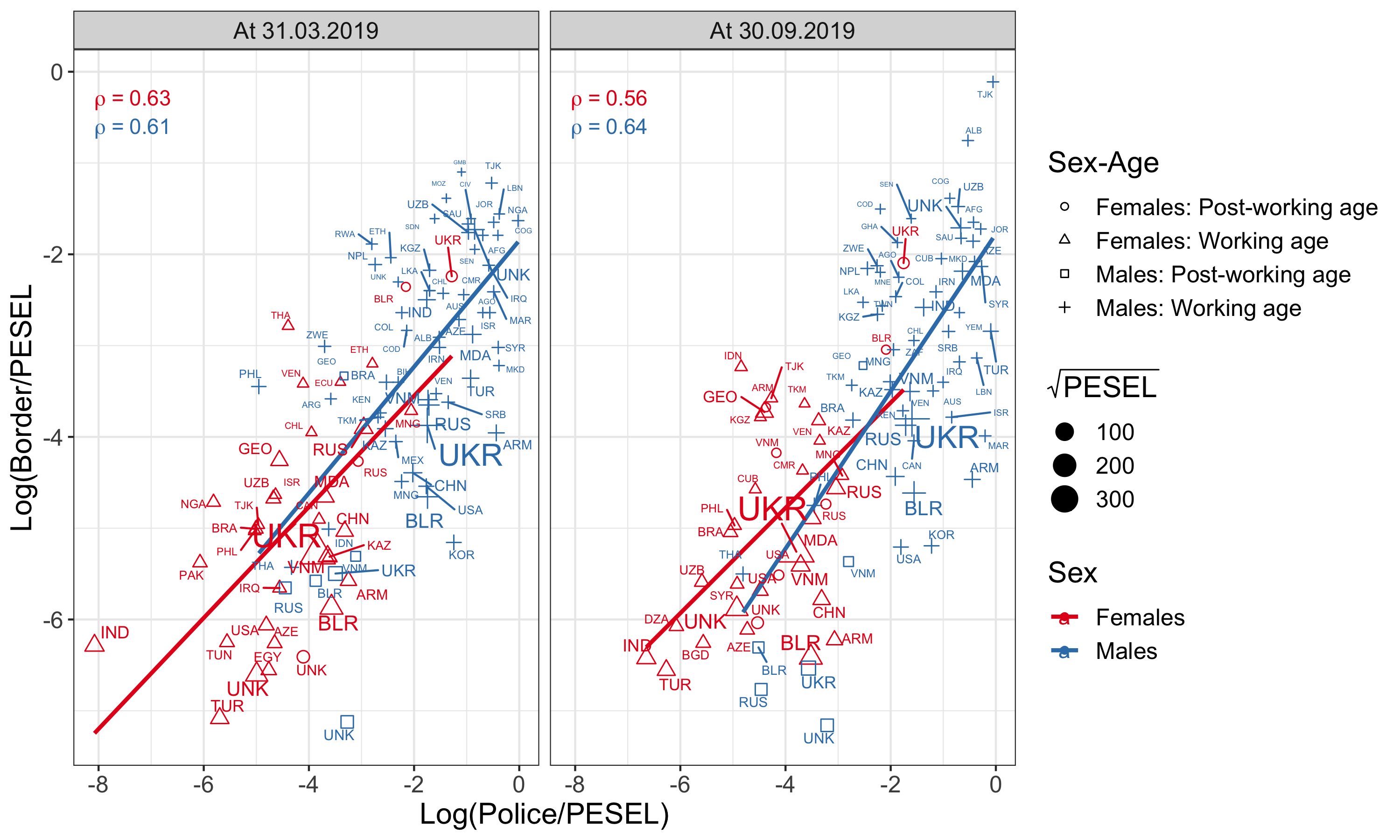}
    \caption{The relationship between the log of the PESEL population and the log of the BG-to-PESEL counts (top) and between the log of police-to-PESEL counts and the log of BG-to-PESEL counts (bottom) at the end of first and third quarter of 2019 by sex. Shapes represent domains cross-classified by sex and age, symbol size represents the square root of the PESEL population and solid lines are regression lines. Pearson correlation coefficient is denoted by $\rho$ in the top left corner.}
    \label{fig-sex}
\end{figure}

To verify the model assumptions we investigate the relationships resulting from equation \eqref{eq-lin} and compare the log of the PESEL population with the log of Border Guard (BG) counts to the PESEL population (top) and the log of police counts to the PESEL population by country of origin (bottom) and sex in both halves of 2019. Figure \ref{fig-overall} presents these relationships with a~linear model defined in \eqref{eq-lin}, which was calculated for the whole dataset, while figure \ref{fig-sex} includes separate fits for each sex. The shapes are defined by the interaction of sex and age (working age and post-working age).

Both plots show the expected relationship, i.e. a~negative correlation with the population size (less than -0.6) and a~positive correlation with the proportion of police-to-PESEL counts for both quarters (over 0.7). This means that the relationship between the size of the unauthorized and registered population decreases as the registered population grows. However, there is an outlier in our population -- Ukraine. Citizens of this country are the biggest immigrant group in Poland in all datasets (over 70\% in the PESEL population, around 60\% of BG apprehensions and close to 70\% of all police registrations). Ukraine is an outlier for both sexes but not for the relationship seen within the police data. If Ukraine is excluded, the correlation with the PESEL population changes to around -0.7 while the correlation with the log of police-to-PESEL counts stays the same. In addition, the pseudo-country, denoted by UNK, is an outlier but only for males.

Figure \ref{fig-sex} presents the same relationship but separately for each sex. As can be seen, there are differences in this respect, particularly in the comparison with the log of the PESEL population, as evidenced by the shift in the regression lines. This means that the unauthorized population mainly consists of males, in contrast to the registered (PESEL) population, which is dominated by females. A similar pattern can be observed regarding the relationship with the police data, where Pearson's correlation coefficient for both sexes is around 0.5-0.6, while without accounting for sex -- around 0.7. 

The above claims are also confirmed by results from fitting the linearized model given by \eqref{eq-lin}. For both periods $\alpha-1$ parameter associated with $\log(N_{tij})$ was equal to -0.4109 and -0.4190 indicating that the relationship with regular population is stable overtime and $\beta$ for $\log(n_{tij}/N_{tij})$ was equal to 0.5694 and 0.5841.

\subsection{Estimation results}

Table \ref{tab-models-comp} contains the main model performance measures, while additional details, including diagnostics, are presented in Appendix \ref{appen-diagnostics}. Results are broken down by quarter end, distribution and covariates used in the modelling phase. We also report AIC and BIC. As expected, truncated distributions yield better lower values of information criteria and higher values of $\hat{\xi}$.

\begin{table}[ht]
\centering
\caption{Quality of models used in the study and the estimated population $\hat{\xi}$}
\label{tab-models-comp}
\begin{tabular}{rrrrrr}
  \hline
Distribution & Covariates for $\alpha$ & LogLik & AIC & BIC & $\hat{\xi}$  \\ 
   \hline
  \multicolumn{6}{c}{At the end of 1$^{st}$ quarter 2019} \\
  \hline
PO & No covariates & -733.1 & 1,470.3 & 1,475.5 & 24,119.9 \\ 
   & Ukraine & -648.7 & 1,303.5 & 1,311.3 & 20,835.8 \\ 
   & Sex  & -682.5 & 1,371.0 & 1,378.8 & 51,982.8 \\ 
   & Ukraine \& Sex & -630.1 & 1,268.1 & 1,278.6 & 34,870.1 \\ 
   \hline
  NB2 & No covariate & -285.7 & 577.4 & 585.2 & 9,664.0 \\ 
   & Ukraine & -283.1 & 574.1 & 584.5 & 11,817.1 \\
   & Sex   & -285.6 & 579.1 & 589.6 & 10,447.2 \\
   & Ukraine \& Sex  & -283.1 & 576.1 & 589.1 & 11,568.0 \\ 
   \hline
  Truncated PO & No covariate & -721.2 & 1,446.4 & 1,451.6 & 24,799.2 \\ 
   & Ukraine & -636.2 & 1,278.4 & 1,286.3 & 21,476.9 \\ 
   & Sex & -657.4 & 1,320.8 & 1,328.6 & 64,142.1 \\
   & Ukraine \& Sex  & -608.9 & 1,225.8 & 1,236.2 & 42,769.8 \\ 
   \hline
  Truncated NB2 & No covariate & -267.1 & 540.2 & 548.0 & 11,390.6 \\ 
   & Ukraine & -264.9 & 537.8 & 548.2 & 14,453.0 \\
   & Sex & -266.4 & 540.8 & 551.2 & 14,239.6 \\
   & Ukraine \& Sex  & -264.7 & 539.5 & 552.5 & 15,959.0 \\ 
  \hline
  \hline
  \multicolumn{6}{c}{At the end of 3$^{rd}$ quarter of 2019} \\
  \hline
  \hline
  PO & No covariate & -822.2 & 1,648.3 & 1,653.4 & 23,582.6 \\
   & Ukraine & -735.7 & 1,477.5 & 1,485.1 & 21,139.0 \\ 
   & Sex   & -742.2 & 1,490.3 & 1,497.9 & 65,011.0 \\
   & Ukraine \& Sex   & -689.8 & 1,387.6 & 1,397.8 & 49,080.1 \\
   \hline
  NB2 & No covariate & -278.8 & 563.6 & 571.2 & 11,421.8 \\ 
   & Ukraine  & -276.5 & 561.1 & 571.3 & 14,568.7 \\ 
   & Sex & -276.6 & 561.2 & 571.4 & 19,128.5 \\
   & Ukraine \& Sex  & -275.5 & 561.0 & 573.7 & 20,258.7 \\ 
   \hline
  Truncated PO & No covariate & -812.9 & 1,629.9 & 1,635.0 & 24,043.0 \\ 
   &  Ukraine & -725.4 & 1,456.7 & 1,464.3 & 21,615.4 \\ 
   &  Sex  & -718.0 & 1,442.0 & 1,449.6 & 80,318.6 \\ 
   & Ukraine \& Sex  & -666.5 & 1,341.1 & 1,351.2 & 61,718.0 \\ 
   \hline
  Truncated NB2 & No covariate & -258.4 & 522.8 & 530.4 & 14,377.1 \\
   & Ukraine  & -256.5 & 521.1 & 531.3 & 19,388.6 \\ 
   & Sex  & -253.8 & 515.6 & 525.8 & 45,008.1 \\ 
   & Ukraine \& Sex & -253.3 & 516.6 & 529.3 & 48,387.7 \\ 
   \hline
\end{tabular}
\end{table}

For both periods, the Poisson and truncated Poisson distributions perform poorly and the estimated irregular population is very large. Results from Table \ref{tab-models-comp} indicate that at the end of the first quarter the truncated NB2 with no covariates or with one covariate for $\alpha$, i.e. Ukraine, is the best model in terms of information criteria (BIC). For the end of the third quarter, the best models also assume the NB2 distribution but the ranking of covariates is different i.e. the model that accounts for sex is the best (BIC=529.3), while the model with Ukraine as a covariate is slightly worse (BIC=531.3). The main difference between these models is the degree of uncertainty, since in the first model the confidence interval is narrower than in the second. There is no justification for such an increase between two periods, given that the regular population grew from 232,500 to 234,200 and a big change in the irregular population is unlikely. This result is mainly due to high values of $\hat{\alpha}_0$, which for the truncated NB2 with sex as a covariate equals 0.875, with sex and Ukraine -- 0.838 and the model with Ukraine -- 0.673. Based on that we decided to focus on truncated NB2 models without covariates and that with Ukraine as the only covariate in $\hat{\alpha}$.

Estimated $\hat{\alpha}_0, \hat{\alpha}_{1}, \hat{\beta}$ and $\hat{\phi}$ are reported in Table \ref{tab-models-estimates}. In addition we provide the Sum of Squares (in thousands) denoted by SSQ. Diagnostics for the final model are presented in Appendix \ref{appen-diagnostics}. For both quarters models with no covariates are characterised by higher $\hat{\alpha}_0$ and $\hat{\beta}$ and the SSQ over 5 times as high as that for the models with one covariate (Ukraine, $\hat{\alpha}_{1}$). As expected, the parameter for Ukraine is positive but is characterised by a high standard error as we have only 4 observations for this country. 

\begin{table}[ht!]
\centering
\caption{Estimated parameters for models with no covariates (no cov.) and with Ukraine as a covariate under the truncated NB2 distribution. Standard errors are reported in parenthesis.}
\label{tab-models-estimates}
\begin{tabular}{lllllll}
  \hline
As at & Model & $\hat{\alpha}_0$ & $\hat{\alpha}_1$ & $\hat{\beta}$ & $\hat{\phi}$ & SSq\\ 
  \hline
31.03 & No cov. & 0.685 (0.032) &  -- & 0.710 (0.067) & 1.267 (0.320) & 823.3\\ 
  31.03 & Ukraine & 0.649 (0.034) & 0.095 (0.05) & 0.665 (0.067) & 1.367 (0.350) & 149.6 \\ 
  31.03 & No cov. & 0.712 (0.038) & -- & 0.814 (0.081) & 0.914 (0.246) &  818.2\\ 
  31.03 & Ukraine & 0.673 (0.041) & 0.104 (0.06) & 0.761 (0.081) & 0.975 (0.263)  & 179.2\\
   \hline
\end{tabular}
\end{table}

The research on irregular migration in Poland is limited. As far as we know, the only results about the unauthorized population in Poland can be found in \cite{pew2019}. The analysis was carried out for the period 2014-2017, and the  population was estimated to be lower than 100,000, regardless of whether or not waiting asylum seekers were included \citet{pew2019}. In their report, \citeauthor{pew2019} does not provide any point estimates or quantify the uncertainty behind this number. Thus, currently there is no other estimate that our results can be compared with\footnote{Note that Eurostat's data presented in the second section are based on Border Guard data and are not included here}.  

To provide some context, we compare our estimates with relevant statistics on migration to Poland reported by the Office for Foreigners for 2019. Table \ref{tab-comparison-results} contains three indicators that can be connected with illegal stays -- negative decisions issued to applications for temporary and permanent stay and decisions about the compulsory return of an individual to their country of origin. A foreigner who has received a negative decision is obliged to leave Poland within 30 days from the date when the decision was issued. If a foreigner does not leave Poland within this period and is apprehended, they are ordered to return. There are multiple reasons why such an order can be issued, such as illegal stay or work or being considered \textit{persona non grata}\footnote{The full list is provided in Appendix \ref{appen-decision}}. The order to return is issued by the commanding officer of the Border Guard unit or the commanding officer of the locally competent Border Guard unit and most of such orders are given to foreigners who exit Poland and were identified as staying illegally (29,072 obligations in Table \ref{tab-comparison-results} and 25,810 in Table \ref{tab-sg-basic}). 

The number of refusals concerning applications for a temporary and permanent stay is close to 36000 and is significantly higher than our estimates. This is mainly because of a variety of reasons for issuing a negative decision (e.g. not meeting requirements for a temporary stay or detention. The full list is given in Appendix \ref{appen-decision}). Our point estimate is lower than the total number of refusals and orders to return, which suggests that the size of the unauthorized population is plausible. 

\begin{table}[ht]
\centering
\caption{Comparison of estimated $\xi$ classified by Ukraine, age group and sex with data from Polish registers}
\label{tab-comparison-results}
\begin{tabular}{rrrrrrr}
  \hline
Period & Total & Ukraine & Working age & Non-working age & Males & Females \\ 
  \hline
  \multicolumn{7}{c}{$\hat{\xi}$} \\
  \hline
31.03.2019 & 14,453 & 9,378 & 13,586 & 867 & 6,492 & 7,961 \\
30.09.2019 & 19,389 & 13,619 & 18,212 & 1,177 & 8,641 & 10,747 \\
  \hline
  \multicolumn{7}{c}{Refused applications for a temporary stay } \\
  \hline
  2019  & 32,835 & 19,685 & -- & -- & 21,623 & 11,212 \\ 
   \hline
   \multicolumn{7}{c}{Refused applications for a permanent stay} \\
   \hline
   2019  & 3,096 & 434 & -- & -- & 1,674 &  1,180\\ 
  \hline
   \multicolumn{7}{c}{Decisions of return to the country of origin} \\
   \hline
   2019  & 29,072 & 21,694 & -- & -- & 20,774 &  8,298\\ 
  \hline
  
\end{tabular}
\end{table}

Table \ref{tab-comparison-results} contains information about the number of irregular residents from Ukraine and by age and sex. The total in comparison to the regular population in Poland in 2019 (37.97 million) is close to 0.04\% on 31 December 2019, and 0.05\% on 30 September 2019 is small and plausible. \cite{pew2019} reports that the irregular population for most countries is lower than 1\%. 

The demographic structure is also probable except for sex. Ukrainians account for over 65\% the irregular population, which is similar to the percentage of refusals or return decisions for Ukrainians. Most of them are people of working age, since their motivation for migrating to Poland is mainly economic. The main problem is the sex structure. Our estimates show that the majority are females, while all other data (apprehensions, refusals, return decisions, etc.) indicate the opposite. The main reason for this result is the structure of the PESEL register, in which the majority (about 60\%) are women. However, if we compare our estimate to the regular foreign population of males and females, we get 6.8\% and 5.8\% respectively, which indicates that males are more likely to be irregular migrants.

Finally, Table \ref{tab-se-results} contains interval estimates for the size of the irregular population using three different measures: \citet{zhang2008developing} plugin interval, the bootstrap calculated using SPIN and the percentile method. The method used by \citet{zhang2008developing} yields a very wide interval ranging from 4,000 to 67,000 at 31 March and from 4,000 to 134,000 at 30 September. The SPIN and quantile method provides similar intervals for the first period suggesting that in the first period the irregular population ranged from 7,000 to 30,000. For the second period SPIN yields a shorter interval between 6,000 and 57,000. The estimated $\sqrt{\widehat{mse}}$ equals 8,259 and 15,904 respectively. The SPIN method is preferred as the bootstrapped $\hat{\xi}^*$ are right-skewed, as shown in Figure \ref{appen-sec-boot}. 

\begin{table}[ht!]
    \centering
    \caption{Estimated size of the irregular population in Poland in 2019 with 95\% interval estimates based on three methods, MSE and RRMSE based on parametric bootstrap}
    \label{tab-se-results}
    \begin{tabular}{llrrr}
         \hline
        Period & Method & Estimate & Lower & Upper\\ 
        \hline
         31.03.2019 & Plug-in &14,453 & 4,696 & 67,404\\
                    & SPIN & 14,453& 6,802 & 29,381\\
                    & Percentile & 14,453 & 7,616 & 30,651\\
                    & $\sqrt{\widehat{mse}}$ & 8,259 & -- & --\\
                    & $\widehat{rrmse}$ & 45\% & -- & --\\
         30.09.2019 & Plug-in &  19,389& 4,836 & 133,792\\
                    & SPIN &  19,389 & 6,011& 47,681\\
                    & Percentile &  19,389& 9,275 & 56,555\\
                    & $\sqrt{\widehat{mse}}$ & 15,904 & -- & --\\
                    & $\widehat{rrmse}$ & 64\% & -- & --\\
         \hline
    \end{tabular}
\end{table}

\section{Discussion}\label{sec-discussion}

In the paper we propose a~different approach to estimate the~hard-to-reach population of irregular foreigners based on a flexible non-linear count regression model. The approach is an alternative to classic capture-recapture methods based on one or multiple sources and the interpretation of results is more intuitive as the irregular population is conditionally dependent on the regular population. Extending the model for additional covariates and zero-truncated distributions makes it more general. That said, the proposed model has certain limitations. 

The approach is based solely on the administrative data and, as a result, the quality of our estimates depends on the availability of high-quality register-based statistics. \citet{beresewicz2019estymacja} provided estimates of the size of the \textit{de facto} population of foreigners for 2015 and 2016. The paper includes information about the co-occurrence of regular foreigners in the PESEL and two external registers maintained by the National Insurance Institution (ZUS) and the Office for Foreigners. For instance, the PESEL register data for 2016 were linked with two other sources and around 7,000 (out of 47,000) foreigners were observed exclusively in the PESEL register. This figure, however, cannot be used as a measure of overcoverage because only three sources were used. In a~more recent study, \citet{gus2020} published a~detailed analysis of foreigners based on 9 registers linked by the PESEL id. Only about 1,500 out of 2.1 million foreigners were found to be listed exclusively in the PESEL register while over 980,000 were listed only in one of the other registers. This indicates that the PESEL register is not considerably affected by overcoverage.

Because not all countries have a population register (e.g. United States or Ireland), it is possible to use population surveys, such as the Current Population Survey conducted by the US Census Bureau, or a system of integrated registers with signs-of-life methodology as in \citet{zhang2018trimmed}. 

Selection of data for the model should be strictly connected with the definition of the irregular population used in the study. Currently, there is no information about how long apprehended foreigners have been staying in Poland, which means that this group can include a mix of persons who have been residing for a period longer than 3 months, have exceeded their temporary residence permit or have been staying without any permit. Therefore, there is a~need for a~close collaboration with the Border Guard.

However, the model assumes a relationship between the regular and irregular population and therefore the approach can be applied to different populations, such as illegal workers or the homeless population, given the existence of register-based proxy populations and auxiliary variables. 

\section*{Acknowledgements}

This work is partially based on Katarzyna Pa\-wlu\-kie\-wicz Master's thesis entitled \textit{Estimation of the number of irregular migrants in Poland using hierarchical Gamma-Poisson model} defended on 19 September 2019 at Poznań University of Economics and Business in Poland.

We thank Ministry of Digital Affairs, Polish Border Guards and Polish Police for compiling the summaries according to our requirements. 

\appendix

\section{List of countries used in the study}\label{appen-data}

\textbf{Our analysis included the following countries}: Afghanistan, Albania, Algeria, Angola, Argentina, Armenia, Australia, Azerbaijan, Bahrain, Bangladesh, Belarus, Bosnia and Herzegovina, Brazil, Cameroon, Canada, Chile, China, Colombia, Congo, Côte D'Ivoire, Cuba, Democratic Republic of the Congo, Ecuador, Egypt, Ethiopia, Gambia, Georgia, Ghana, India, Indonesia, Iran, Iraq, Israel, Jordan, Kazakhstan, Kenya, Kosovo, Kyrgyzstan, Lebanon, Mexico, Mongolia, Montenegro, Morocco, Mozambique, Nepal, New Zealand, Nigeria, North Macedonia, Pakistan, Philippines, Republic of Korea, Republic of Moldova, Russian Federation, Rwanda, Saudi Arabia, Senegal, Serbia, South Africa, Sri Lanka, Sudan, Syrian Arab Republic, Taiwan, Tajikistan, Thailand, Tunisia, Turkey, Turkmenistan, Ukraine, United States of America, Uzbekistan, Venezuela, Vietnam, Yemen, Zimbabwe and other.

\textbf{Note:} In the second half of 2019, no women of post-working age (60+) were apprehended by the Border Guard, while the PESEL register contained 1,286 foreigners and police records -- 52. As we did not have any other group to merge with we decided to artificially add one person to the Border Guard dataset to represent this subgroup in that period. The same number of foreigners in this subgroup were apprehended in the first half year.

\section{Log-likelihood functions, derivatives and simulation study}\label{appen-ll-der}

\subsection{The estimation procedure of \citet{zhang2008developing} model}\label{appen-estimation-mixture}

\citet{zhang2008developing} used the maximum likelihood method to estimate $\alpha$ and $\beta$. Let $L(\boldeta, \boldm)$ denote the likelihood of $\boldeta = (\alpha, \beta, \phi)$ given $m_i$, for $i=1,...,C$. To simplify the notation, we will use $\mu_i$ for $(\alpha, \beta)$ parameters as they are derived from  $\mu_i=N_i^\alpha(n_i/N_i)^\beta$. Under the Poisson gamma model \eqref{eq-zhang-model}, we have

\begin{equation}
f \left( m _ { i } , u _ { i } ; \boldeta \right) = \frac { e ^ { - \mu _ { i } u _ { i } } \left( \mu _ { i } u _ { i } \right) ^ { m _ { i } } } { m _ { i } ! } \cdot \frac { \phi ^ { \phi } u _ { i } ^ { \phi - 1 } e ^ { - \phi u _ { i } } } { \Gamma ( \phi ) } = \frac { \mu _ { i } ^ { m _ { i } } \phi ^ { \phi } } { m _ { i } ! \Gamma ( \phi ) } e ^ { - u _ { i } \left( \mu _ { i } + \phi \right) } u _ { i } ^ { m _ { i } + \phi - 1 },
\end{equation}

\noindent where $\Gamma()$ is the gamma function. Thus,

\begin{equation}
\begin{aligned} f \left( m _ { i } ; \boldeta \right) & = \int _ { 0 } ^ { \infty } f \left( m _ { i } , u _ { i } ; \boldeta \right) d \left( u _ { i } \right) \\ & = \frac { \mu _ { i } ^ { m _ { i } } \phi ^ { \phi } } { m _ { i } ! \Gamma ( \phi ) } \int _ { 0 } ^ { \infty } e ^ { - \left( \sqrt { u _ { i } } \right) ^ { 2 } \left( \mu _ { i } + \phi \right) } \left( \sqrt { u _ { i } } \right) ^ { 2 \left( m _ { i } + \phi - 1 \right) } 2 \sqrt { u _ { i } } d \left( \sqrt { u _ { i } } \right) \\ & = \frac { \mu _ { i } ^ { m _ { i } } \phi ^ { \phi } } { m _ { i } ! \Gamma ( \phi ) } \left( \mu _ { i } + \phi \right) ^ { - \left( m _ { i } + \phi \right) } \Gamma \left( m _ { i } + \phi \right),\end{aligned}
\label{eq-pois-gamma-density}
\end{equation}

\noindent based on the identity $\int _ { 0 } ^ { \infty } e ^ { - \gamma z ^ { 2 } } z ^ { k } d z = \frac { 1 } { 2 } \gamma ^ { - \frac { k + 1 } { 2 } } \Gamma \left( \frac { k + 1 } { 2 } \right),$ with $z=\sqrt{u_i}$ and $k=2(m_i + \phi) - 1$. Conditional on $m_i$, $u_i$ has the gamma distribution with mean $(m_i + \phi)/(\mu_i + \phi)$ and variance $(m_i + \phi) / (\mu_i + \phi)^2$.

As noted by \citet{cameron2013regression}, the representation of the negative binomial distribution as a~Poisson-Gamma mixture is an old result following back to \citet{greenwood1920inquiry} and the parametrization of the gamma function leads to different variance functions \citep{cameron1986econometric}. 

Now, based on \citet[pp. 117-118]{cameron2013regression}, we show that \eqref{eq-pois-gamma-density} is actually a negative binomial distribution represented as Gamma-Poisson mixture. Let $\kappa=1/ \phi$ and from the gamma function property $\Gamma(m+1)=m!$ we get 

\begin{equation}
    \begin{aligned}
    f \left( m _ { i } ; \boldeta \right) & = 
    \frac { \mu _ { i } ^ { m _ { i } } (\kappa^{-1}) ^ { \kappa^{-1} } } { m _ { i } ! \Gamma ( \kappa^{-1} ) } \left( \mu _ { i } + \kappa^{-1} \right) ^ { - \left( m _ { i } + \kappa^{-1} \right) } \Gamma \left( m _ { i } + \kappa^{-1} \right) \\
    & = \frac{\Gamma(m_i+\kappa^{-1})}{\Gamma(\kappa^{-1})\Gamma(m_i+1)}\mu^{m_i}(\kappa^{-1})^{\kappa^{-1}}(\mu+\kappa^{-1})^{-(m_i+\kappa^{-1})} \\
    & = \frac{\Gamma(m_i+\kappa^{-1})}{\Gamma(\kappa^{-1})\Gamma(m_i+1)}\mu^{m_i}(\mu+\kappa^{-1})^{-m_i} (\kappa^{-1})^{\kappa^{-1}}(\mu+\kappa^{-1})^{-\kappa^{-1}} \\
    & = \frac{\Gamma(m_i+\kappa^{-1})}{\Gamma(\kappa^{-1})\Gamma(m_i+1)} 
    \left(
    \frac{\mu}{\mu+\kappa^{-1}}
    \right)^m
    \left(
    \frac{\kappa^{-1}}{\mu+\kappa^{-1}}
    \right)^{\kappa^{-1}}  \\
    & = \frac{\Gamma(m_i+ 1/\phi)}{\Gamma(1/ \phi)\Gamma(m_i+1)} 
    \left(
    \frac{\mu}{\mu+ 1/\phi}
    \right)^m
    \left(
    \frac{1 / \phi}{\mu+ 1 / \phi }
    \right)^{1 / \phi},
    \end{aligned}
\end{equation}

\noindent which is an alternative parameterisation of Negative Binomial with mean $\mu$ and dispersion parameter $\phi$. Another way to specify the Gamma-Poisson mixture as a Negative Binomial is given in \citet[p. 117-118]{cameron2013regression} .

The likelihood using \eqref{eq-pois-gamma-density} is given by

\begin{equation}
L ( \boldeta ; \mathbf { m } ) = \prod _ { i = 1 } ^ { C } f \left( m _ { i } ; \boldeta \right),
\end{equation}

\noindent Thus, disregarding constant terms, the log-likelihood is given by

\begin{equation}
    l ( \boldeta ; \mathbf { m } ) = \sum _ { i = 1 } ^ { C} l _ { i } ( \boldeta ),
\end{equation}

\noindent where \citet{zhang2008developing} proposed to used the following log-likelihood function

\begin{equation}
    \begin{aligned} 
    l _ { i } ( \boldeta ) & = m _ { i } \log \mu _ { i } - \left( m _ { i } + \phi \right) \log \left( \mu _ { i } + \phi \right) + \log \Gamma \left( m _ { i } + \phi \right) + \phi \log \phi - \log \Gamma ( \phi ) \\ 
    & \doteq m _ { i } \log \mu _ { i } - \left( m _ { i } + \phi \right) \log \left( \mu _ { i } + \phi \right) + \phi \log \phi \\ 
    & + \left( m _ { i } + \phi - 0.5 \right) \log \left( m _ { i } + \phi \right) - \left( m _ { i } + \phi \right) - ( \phi - 0.5 ) \log ( \phi ) + \phi  \\ 
    & = m _ { i } \log \mu _ { i } - \left( m _ { i } + \phi \right) \log \left( \mu _ { i } + \phi \right) + \left( m _ { i } + \phi - 0.5 \right) \log \left( m _ { i } + \phi \right) + 0.5 \log \phi, 
    \label{eq-ll-zhang}
    \end{aligned}
\end{equation}

\noindent where $\mu_i$ is defined as in \eqref{eq-mu} and \citet{zhang2008developing} used the Stirling's approximation given by $\log \Gamma ( z ) \doteq ( z - 0.5 ) \log ( z ) + 0.5 \log ( 2 \pi ) - z$. 

However, the exact Stirling's approximation is $\log \Gamma ( z ) = ( z - 0.5 ) \log ( z ) + 0.5 \log ( 2 \pi ) - z + \int_{0}^{\infty} \frac{2 \arctan \left(\frac{t}{z}\right)}{e^{2 \pi t}-1} \mathrm{d} t$ and thus the log-likelihood function given in \eqref{eq-ll-zhang} is not complete and may result in biased or inefficient estimates of $\boldeta$. To verify this we conducted a~limited simulation study involving a simplified version proposed by \citet{zhang2008developing} and compared with the exact version mentioned above. The results presented in Appendix \ref{appen-stirling} show that \citet{zhang2008developing} proposal leads to slight bias in $\alpha, \beta$ and $\xi$ estimates. 

The MLE of $\boldeta$, denoted by $\hat{\boldeta}$, is given by the solution to the likelihood equations, i.e. 

\begin{equation}
    \frac { \partial l ( \boldeta ; \boldm ) } { \partial \boldeta } = \sum _ { i = 1 } ^ { C } \frac { \partial l _ { i } ( \boldeta ) } { \partial \boldeta } = 0.    
\end{equation}

The MLE can be obtained using the Newton-Raphson method with starting values obtained from the model given in \eqref{eq-lin}. In this case, $\alpha$ and $\beta$ are used as the starting values for the same parameters of the Poisson-gamma model, and the inverse of the estimated $V(\epsilon_i)$ is used as the starting value for $\phi$. We estimated the model using \texttt{maxLik} package \citep{maxlik} in the R language \citep{rcran}.

\subsection{Partial derivatives for \cite{zhang2008developing} model}

The mean parameter $\mu_i$ is linear on the log scale, denoted by $log \mu_i = x_i^T\gamma$
with generic vector of covariates $x_i$ and parameters $\gamma$. Now that $l_i(\eta)$ depends on $\gamma$ only through $\mu_i$, we have

\begin{equation}
    \frac { \partial l _ { i } ( \eta ) } { \partial \gamma } = \frac { \partial l _ { i } ( \eta ) } { \partial \mu _ { i } } \frac { \partial \mu _ { i } } { \partial \log \mu _ { i } } \frac { \partial \log \mu _ { i } } { \partial \gamma } = \frac { \partial l _ { i } ( \eta ) } { \partial \mu _ { i } } \mu _ { i } x _ { i } = \frac { m _ { i } - \mu _ { i } } { \mu _ { i } + \phi } \phi x _ { i },
\end{equation}

where $\partial l _ { i } ( \eta ) / \partial \mu _ { i } = m _ { i } / \mu _ { i } - \left( m _ { i } + \phi \right) / \left( \mu _ { i } + \phi \right)$, and 

\begin{equation}
    \frac { \partial l _ { i } ( \eta ) } { \partial \phi } = - \log \left( \mu _ { i } + \phi \right) - \frac { m _ { i } + \phi } { \mu _ { i } + \phi } + \log \left( m _ { i } + \phi \right) + \frac { m _ { i } + \phi - 0.5 } { m _ { i } + \phi } + \frac { 1 } { 2 \phi }.
\end{equation}

Moreover,

\begin{equation}
  \frac { \partial ^ { 2 } l _ { i } ( \eta ) } { \partial \gamma \partial \gamma ^ { T } } = \frac { \partial ^ { 2 } l _ { i } ( \eta ) } { \partial \mu _ { i } ^ { 2 } } \mu _ { i } x _ { i } \frac { \partial \mu _ { i } } { \partial \gamma ^ { T } } + \frac { \partial l _ { i } ( \eta ) } { \partial \mu _ { i } } x _ { i } \frac { \partial \mu _ { i } } { \partial \gamma ^ { T } } = - \left( \frac { m _ { i } + \phi } { \mu _ { i } + \phi } \phi \right) x _ { i } x _ { i } ^ { T },  
\end{equation}

\begin{equation}
    \frac { \partial ^ { 2 } l _ { i } ( \eta ) } { \partial \phi ^ { 2 } } = - \frac { 2 \mu _ { i } + \phi - m _ { i } } { \left( \mu _ { i } + \phi \right) ^ { 2 } } + \frac { m _ { i } + \phi + 0.5 } { \left( m _ { i } + \phi \right) ^ { 2 } } - \frac { 1 } { 2 \phi ^ { 2 } },
\end{equation}

\noindent and 

\begin{equation}
    \frac { \partial ^ { 2 } l _ { i } ( \eta ) } { \partial \gamma \partial \phi } = \left( \partial \left( \frac { \partial l _ { i } ( \eta ) } { \partial \mu _ { i } } \right) / \partial \phi \right) \mu _ { i } x _ { i } = - \frac { \mu _ { i } - m _ { i } } { \left( \mu _ { i } + \phi \right) ^ { 2 } } \mu _ { i } x _ { i } = \left( \frac { \partial ^ { 2 } l _ { i } ( \eta ) } { \partial \phi \partial \gamma ^ { T } } \right) ^ { T }.
\end{equation}

\subsection{Likelihood functions for models used in the study}

\subsubsection{Poisson (PO)}

\begin{equation}
\begin{aligned}
    l(\boldeta) = l(\boldalpha, \beta) & = m_i \log(\mu_i) - \mu_i  - log(m_i!).
\end{aligned}
\end{equation}

\subsubsection{zero-truncated Poisson (ztPO)}

\begin{equation}
\begin{aligned}
    l(\boldeta) = l(\boldalpha, \beta) & = m_i \log(\mu_i) - \mu_i - \log(m_i!) - \log(1 - e^{-\mu_i}).
\end{aligned}
\end{equation}

\subsubsection{Negative Binomial type 2 (NB2)}

\begin{equation}
\begin{aligned}
    l(\boldeta) = l(\boldalpha, \beta, \phi) & =  \sum_{v=0}^{m_{i}} \log \left(v+\phi\right)-\log m_{i}! -\left(m_{i}+\phi\right) \log \left(1+ \frac{\mu_{i}}{\phi} \right) \\
    & + m_{i} \log \frac{\mu_{i}}{\phi}.
\end{aligned}
\end{equation}

\subsubsection{zero-truncated Negative Binomial type 2 (ztNB2)}

\begin{equation}
\begin{aligned}
    l(\boldeta) = l(\boldalpha, \beta, \phi) & =  \sum_{v=0}^{m_{i}} \log \left(v+\phi\right)-\log m_{i}! -\left(m_{i}+\phi\right) \log \left(1+ \frac{\mu_{i}}{\phi} \right) \\
    & + m_{i} \log \frac{\mu_{i}}{\phi}-\log \left[1-\left(1+\frac{\mu_{i}}{\phi}\right)^{-\phi}\right].
\end{aligned}
\end{equation}

\subsection{Limited simulation study -- Stirling's approximation, zero-truncation and its impact on estimates}\label{appen-stirling}

\cite{zhang2008developing} used Stirling's approximation of $\log\Gamma(x)$ that drops the integral part and the log-likelihood function in \eqref{eq-ll-zhang} is reduced. To verify whether this approach has an impact on the estimated parameters we conducted a~limited simulation study using Polish data in which we compared log-likelihood using \eqref{eq-ll-zhang} (denoted as \textit{zhang}), log-likelihood using \eqref{eq-ll-zhang} but $\log\Gamma(x)$ was calculated using the \texttt{lgamma} function in R (denoted as \textit{lgamma}), log-likelihood using the \texttt{dnbinom} function in R (denoted as \textit{dnbinom}) and log-likelihood for zero-truncated NB2 using \texttt{dztnbinom} from the \texttt{actuar} package in R (denoted as \textit{dztnbinom}).

We set $\alpha=0.7, \beta=0.8$ and considered two cases: when the dispersion parameter is low ($\phi=1.5$) and when it is high ($\phi=2.5$). We generated $m_i$ data using \texttt{rnbinom} with mean parametrisation, where $\mu_i$ were defined as in \eqref{eq-mu}. We used $N_i$ and $n_i$ for the first quarter of 2019. In each iteration (B=500) we removed cases when $m=0$ to mimic Border Guard data. To assess the performance we report the following measures:

\begin{itemize}
    \item Relative bias
    \begin{equation}
    \text{RB}_{\text{SIM}}(\hat{\theta}) = \frac{1}{B}\sum_{b=1}^B\frac{\hat{\theta}^{(b)}-\theta}{\theta} \times 100\%,
\end{equation}
    \item Relative root m square error
    \begin{equation}
        \text{RRMSE}_{\text{SIM}}(\hat{\theta}) = \frac{\sqrt{\text{MSE}_{\text{SIM}}}}{\theta} \times  100\%,
    \end{equation}
    where $\text{MSE}_{\text{SIM}}$ is calculated as 
    \begin{equation}
    \text{MSE}_{\text{SIM}}(\hat{\theta}) = \frac{1}{B}\sum_{b=1}^B\left(\hat{\theta}^{(b)} - \theta \right)^2
\end{equation}
\end{itemize}

\noindent where $\theta$ is replaced with $\alpha, \beta, \phi$ and $\xi$. Results are presented in Table \ref{tab-sim-results}. The approximation used by Zhang leads to biased estimates of $\alpha$, $\beta$ and $\xi$, but for $\phi$ in both cases (low and high) the relative bias is lower than that obtained using the \texttt{lgamma} or \texttt{dnbinom} function. This means that \eqref{eq-ll-zhang} should not be used for estimating the \eqref{eq-zhang-model} model. Furthermore, mis-specification of the distribution (NB2 instead of zero-truncated NB2) leads to underestimation of the size of the irregular population. 

\begin{table}[ht!]
\centering
\caption{Comparison of estimated parameters using different specifications of log-likelihood functions based on simulation study}
\label{tab-sim-results}
\begin{tabular}{lllrr}
  \hline
Dispersion & Parameter & Log-Lik & $\text{RB}_{\text{SIM}}$ & $\text{RRMSE}_{\text{SIM}}$ \\ 
  \hline
  High ($\phi=2.5$) & $\alpha$ & \textit{dnbinom} & -3.98 & 1,990.05 \\ 
   &  & \textit{lgamma} & -4.01 & 2,003.49 \\ 
   &  & \textit{zhang} & -4.06 & 2,029.44 \\ 
   &  & \textit{ztbinom} & -0.53 & 264.68 \\ 
   & $\beta$ & \textit{dnbinom} & -13.01 & 6,505.62 \\ 
   &  & \textit{lgamma} & -13.03 & 6,516.29 \\ 
   &  & \textit{zhang} & -13.25 & 6,623.00 \\ 
   &  & \textit{ztbinom} & -0.70 & 349.89 \\ 
   & $\phi$ & \textit{dnbinom} & 126.95 & 63,474.46 \\ 
   &  & \textit{lgamma} & 125.05 & 62,524.28 \\ 
   &  & \textit{zhang} & 108.77 & 54,384.40 \\ 
   &  & \textit{ztbinom} & 82.04 & 41,021.30 \\ 
   & $\xi$ & \textit{dnbinom} & -12.39 & 6,192.63 \\ 
   &  & \textit{lgamma} & -12.55 & 6,273.28 \\ 
   &  & \textit{zhang} & -12.84 & 6,421.16 \\ 
   &  & \textit{ztbinom}& 9.67 & 4,836.92 \\ 
  Low ($\phi=1.5$) &  $\alpha$  & \textit{dnbinom} & -4.56 & 2,278.80 \\ 
   &  & \textit{lgamma} & -4.61 & 2,303.05 \\ 
   &  & \textit{zhang} & -4.64 & 2,319.28 \\ 
   &  & \textit{ztbinom} & -0.52 & 257.70 \\ 
   & $\beta$ & \textit{dnbinom} & -15.86 & 7,932.38 \\ 
   &  & \textit{lgama}& -15.97 & 7,983.36 \\ 
   &  & \textit{zhang} & -16.13 & 8,063.56 \\ 
   &  & \textit{ztbinom} & -0.36 & 181.96 \\ 
   & $\phi$ & \textit{dnbinom} & 50.17 & 25,083.16 \\ 
   &  & \textit{lgamma} & 51.01 & 25,504.54 \\ 
   &  & \textit{zhang} & 35.35 & 17,676.58 \\ 
   &  & \textit{ztbinom} & 9.07 & 4,532.78 \\ 
   & $\xi$ & \textit{dnbinom} & -15.09 & 7,545.81 \\ 
   &  & \textit{lgamma} & -15.36 & 7,680.53 \\ 
   &  & \textit{zhang} & -15.56 & 7,780.85 \\ 
   &  & \textit{ztbinom} & 11.15 & 5,575.33 \\ 
   \hline
\end{tabular}
\end{table}

\section{Diagnostics for the ztNB2 model}\label{appen-diagnostics}

\subsection{Residuals}

\citet[ch. 5.2.1]{cameron2013regression} discuss Pearson, Deviance and Anscombe residuals to assess the quality of the model. In the paper we focus on the latter, which for the Negative Binomial is given by

\begin{equation}
r_{i}=\frac{
\frac{3}{\kappa}
\left\{\left(1+\kappa m_i\right)^{2 / 3}-\left(1+\kappa \hat{\mu}_{i}\right)^{2 / 3}\right\}
+3\left(m_i^{2 / 3}-\hat{\mu}_{i}^{2 / 3}\right)}
{2\left(\hat{\mu}_{i}+\kappa \hat{\mu}_{i}^{2}\right)^{1 / 6}},
\label{eq-anscombe}
\end{equation}

\noindent where $\kappa = 1/ \phi$.

\begin{figure}[ht!]
    \centering
    \includegraphics[width=0.7\textwidth]{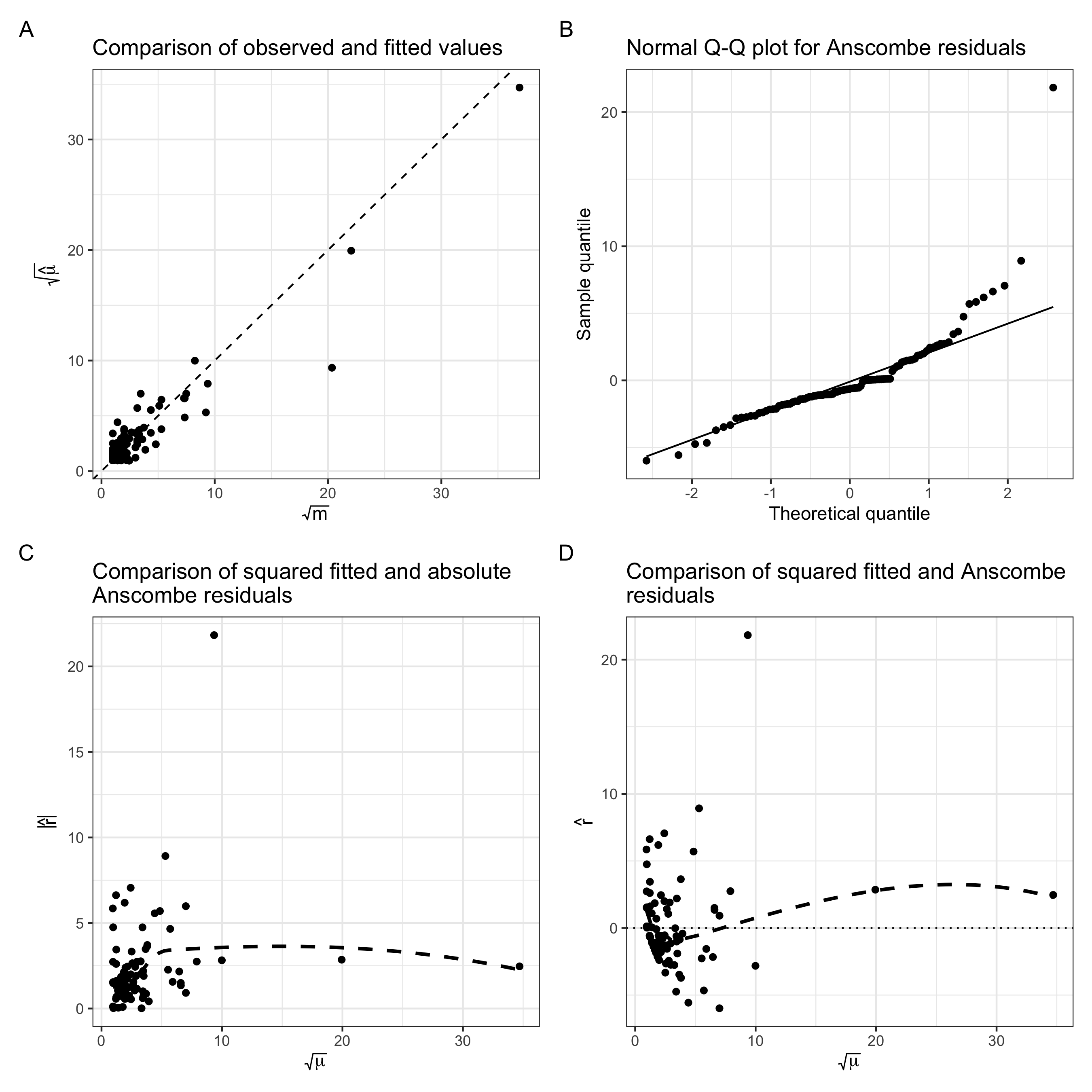}
    \caption{Diagnostics for the model for 31 Mar 2019}
    \label{fig-resid-1}
\end{figure}

\begin{figure}[ht!]
    \centering
    \includegraphics[width=0.7\textwidth]{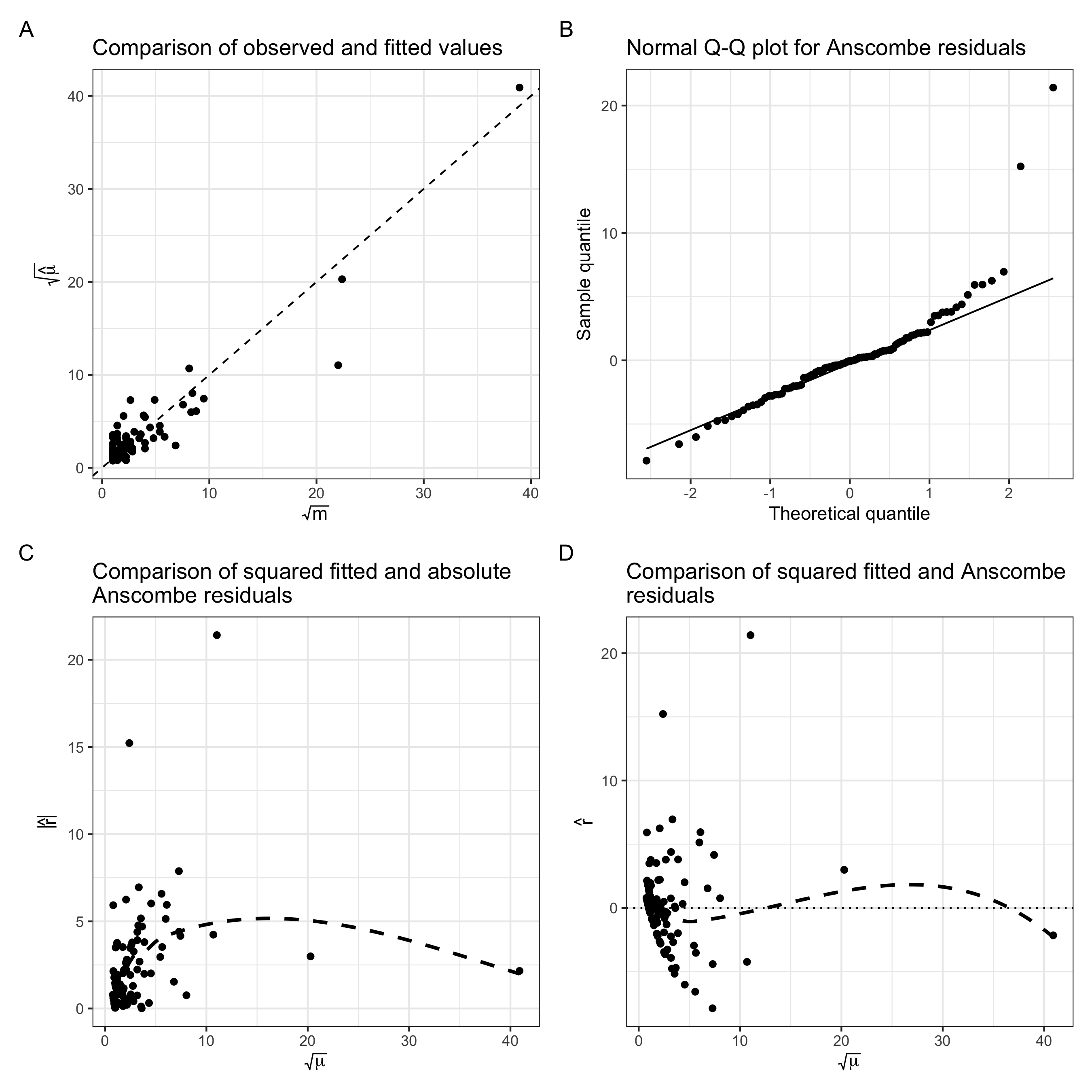}
    \caption{Diagnostics for the model for 30 Sept 2019}
    \label{fig-resid-2}
\end{figure}

Figure \ref{fig-resid-1} and \ref{fig-resid-2} present diagnostic plots for the first and the second model. Both plots contain four subplots showing comparisons of observed, fitted and Anscombe residuals calculated using \eqref{eq-anscombe} formulae. Residuals are skewed because of poorly fitted values for males of working age from the pseudo-country \textit{rest} (414 vs 87.38), India (85 vs 28.12) and the Philippines (9 vs 1.45), and females of working age from Georgia (23 vs 5.86) in the first quarter. In the case of the second model, there were problems with males of working age from the pseudo-country \textit{rest} (485 vs 121.69), Armenia (7 vs 53.05) and Tajikistan (34 vs 11.08) and females of working age from Georgia (47 vs 5.74).

\subsection{Bootstrap}\label{appen-sec-boot}

\begin{figure}[ht!]
    \centering
    \includegraphics[width=0.9\textwidth]{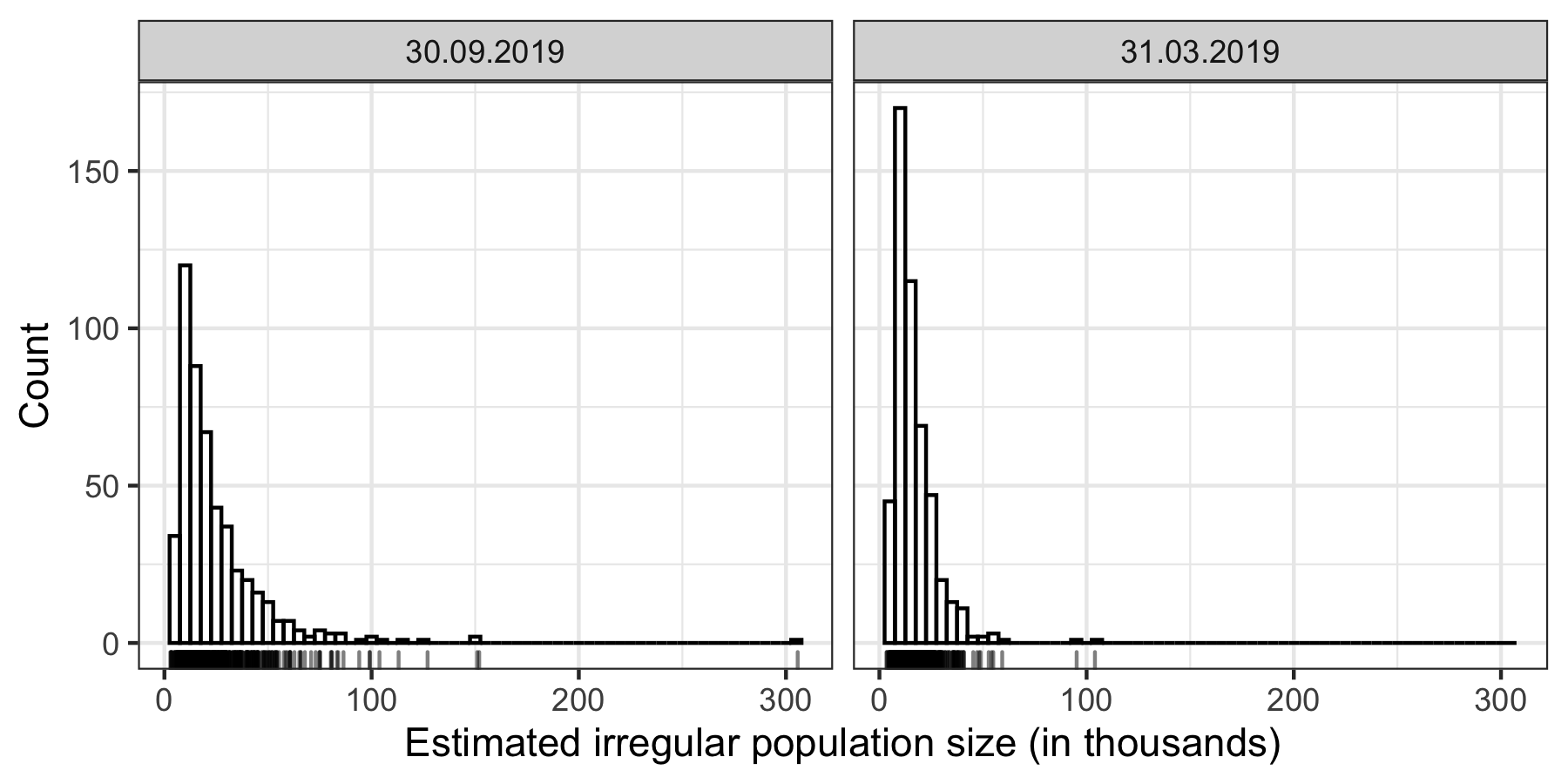}
    \caption{Distribution of the bootstrapped estimates of the irregular population size ($\hat{\xi}^*$)}
    \label{fig-boot-mse}
\end{figure}

Figure \ref{fig-boot-mse} presents a  distribution of bootstrapped estimates of the irregular population size ($\hat{\xi}^*$). For both periods, distributions are highly skewed and the SPIN method is recommended. The mean and median for the first quarter are 16,563 and 13,826, while for the second -- 24,822 and 18,074.

\clearpage

\section{Definitions of measures used for the comparison}

Description below is taken from the official web page of the Office for Foreigners (links: \href{https://udsc.gov.pl/en/cudzoziemcy/obywatele-panstw-trzecich/chce-przedluzyc-swoj-pobyt-w-polsce/zezwolenie-na-pobyt-czasowy/basic-information/}{Temporary residence permit}, \href{https://udsc.gov.pl/en/cudzoziemcy/obywatele-panstw-trzecich/chce-osiedlic-sie-w-polsce/zezwolenie-na-pobyt-staly/basic-information/}{Permanent residence permit}).

\subsection{The decision about the compulsory return of an individual to their country of origin}\label{appen-decision}

A return decision is issued when a foreigner:

\begin{itemize}
    \item resides or resided within the territory of the Republic of Poland without a~valid visa or another valid document authorising him/her to enter this territory and stay within it, if a~visa or the other document is or were required or
    \item has not left the territory of the Republic of Poland after the lapse of the maximum duration of his/her stay within the territory of some or all Schengen countries to which he/she was entitled without the need for a~visa for 180 days in each period, unless international agreements provide otherwise, or
    \item has not left the territory of the Republic of Poland after the lapse using the maximum duration of his/her stay indicated in the Schengen visa within each 180-day period or after the lapse of the permissible period of stay on the basis of a~national visa, or
    \item performs or performed work without a~required work permit or an employer’s declaration of intention (registered in a~district labour office) to employ him/her to perform work, or has been fined for illegal performance of work, or
    \item operated an economic activity in breach of the regulations applicable in this regard within the territory of the Republic of Poland, or
    \item does not have the financial resources necessary to cover the costs of his/her stay within the territory of the Republic of Poland, to travel back to the country of origin or residence or transit through the territory of the Republic of Poland to a~third country that will grant a~permission to enter and has not indicated reliable sources to obtain such funds, or
    \item the foreigner is entered in the register of foreigners whose stay within the territory of the Republic of Poland is undesirable, or
    \item the foreigner’s data can be found in the Schengen Information System for the purposes of refusing entry if the foreigner stays within the territory of the Republic of Poland under the visa-free travel regime or under a~Schengen visa, with the exception of a~visa authorising only the entry and stay within the territory of the Republic of Poland, or
    \item it is justified by national security or defence, the protection of public order and safety or the interests of the Republic of Poland, or
    \item has crossed or attempted to cross the border in breach of legal regulations, or
    \item has been convicted in the Republic of Poland by a~final decision for a~custodial sentence subject to execution, and there are grounds to conduct proceedings on his/her transfer abroad for the purpose of enforcing the penalty against him, or
    \item resides outside the border zone in which according to the permit for crossing the border under the local border traffic, he/she may reside, unless international agreements stipulate otherwise, or
    \item stays within the territory of the Republic of Poland after lapse of the period of stay to which he/she was entitled under a~permit to cross the border under the local border traffic, unless international agreements stipulate otherwise, or
    \item further stay of the foreigner within the territory of the Republic of Poland would be a~threat to public health, which was confirmed by clinical examination, or to the international relations of another European Union Member State, or
    \item the purpose and conditions of stay of a~foreigner within the territory of the Republic of Poland are inconsistent with the declared ones, unless the legal regulations allow him/her to be changed, or
    \item  a~decision on refusal to grant refugee status or subsidiary protection award or a~decision to discontinue the proceedings on granting him/her refugee status was issued and he/she has not left the territory of the Republic of Poland within the deadline and in the case referred to in Article 299(6)(2) of Act of 12 December 2013 on Foreigners.
\end{itemize}

\subsection{Negative decisions for temporary or permanent stay}

Foreigner shall be refused to grant \textbf{temporary residence} permit, if:

\begin{itemize}
    \item that person does not meet the requirements of granting temporary residence permit due to the declared purpose of the stay or the circumstances, which constitute the basis for application for the permit, do not justify his/her stay within the territory of the Republic of Poland for a~period longer than 3 months, or
    \item there is a~valid entry of the foreigner to the register for foreigners, whose stay within the territory of the Republic of Poland is undesirable, or
    \item that persons data are present in the Schengen Information System for the purposes of entry refusal, or
    \item it is required by the State defence and security or maintenance of public order and security or obligations resulting from ratification provisions of the international agreements applying to the Republic of Poland, or
    \item in the proceedings for granting temporary residence permit: a) that persons has made an application containing untrue personal data or false information or attached it with documents containing such data or information, or b) that person has testified untruthfully or has concealed the truth or counterfeited or forged a~document in order to use it as the original or has used such a~document as a~original, or
    \item that persons is in arrears of taxes, with exception of the cases when he/she has obtained a~legal exemption, deferral, division of overdue amounts into instalments or the enforcement of the whole decision of a~competent body has been suspended, or
    \item that person has not returned the costs related with issuance and execution of the decision on obligation of the foreigner to return, which were covered from the State budget, or
    \item being subject to the obligation of treatment pursuant to Article 40 (1) of the Act of 5 December 2008, on  preventing and combating infections and infectious diseases among people he/she refuses to consent to treatment, or
    \item that persons has made an application during an illegal stay within the territory of the Republic of Poland or is illegally staying within the territory.
\end{itemize}

Initiation of proceedings on granting a~\textbf{permanent residence permit} to the foreigner is refused when on the day of submitting the application for such permit the foreigner:

\begin{itemize}
    \item Residing in the territory of the Republic of Poland:
    \begin{itemize}
        \item Illegally or
        \item On the basis of a~Schengen visa authorizing only to enter the territory of the Republic of Poland, and stay in this territory that was issued for the purpose of entry for humanitarian reasons, due to the state's interest or international obligations or
        \item On the basis of a~temporary residence permit due to circumstances requiring a~short stay or
        \item On the basis of a~EU long – term resident residence permit or
    \end{itemize}
    \item Is detained, placed in a~guarded centre, or in detention for foreigners, or a~preventive measure is applied to him/her in the form of prohibition to leave the country or
    \item Is serving prison term or is temporarily detained or
    \item Stays in the territory of the Republic of Poland after he/she has been obliged to return, and the period for voluntary return specified in the decision obliging the foreigner to return has not yet expired, also in the case of extension of this deadline or
    \item Is obliged to leave the territory of the Republic of Poland in the event of refusal to grant or withdrawal of residence permit or in the event of refusal to grant or withdrawal of international protection or
    \item Stays outside the borders of the Republic of Poland.
\end{itemize}

% AOS,AOAS: If there are supplements please fill:

% \begin{supplement}[id=suppA]
%   \sname{Supplement A}
%   \stitle{Log-likelihoods}
%   \slink[doi]{10.1214/00-AOASXXXXSUPP}
%   \sdatatype{.pdf}" 
%   \sdescription{Log-likelihoods and derivatives for models used in the paper}
% \end{supplement}

\begin{supplement}[id=suppA]
  \sname{Supplement A}
  \stitle{R codes}
  \slink[doi]{10.1214/00-AOASXXXXSUPP}
  \sdatatype{.pdf}" 
  \sdescription{R codes to reproduce all results in the paper}
\end{supplement}

\bibliographystyle{imsart-nameyear}
\bibliography{bibliography}

\clearpage

\end{document}